\documentclass{aastex61}
\usepackage{graphicx}
\usepackage{amsmath}
\usepackage[flushleft]{threeparttable}
\usepackage{color}
\usepackage{float}
\usepackage[caption = false]{subfig}

\newcommand{\Hb}{\ensuremath{{\rm H}\beta}}
\newcommand{\Ha}{\ensuremath{{\rm H}\alpha}}

\newcommand{\beq}{\begin{equation}}
\newcommand{\eeq}{\end{equation}}

\newcommand{\kms}{ km s$^{-1}$ }

\shorttitle{All-Sky Optical AGN Catalog}
\shortauthors{Chen et al.}

\begin{document}


\title{A Uniformly Selected, Southern-Sky 6dF, Optical AGN Catalog}

\correspondingauthor{Ingyin Zaw}
\email{iz6@nyu.edu}

\author[0000-0001-8821-0309]{Yan-Ping Chen}
\affiliation{New York University Abu Dhabi,
    P.O. Box 129188, Abu Dhabi, United Arab Emirates}

\author[0000-0002-5208-1426]{Ingyin Zaw}
\affil{New York University Abu Dhabi,
    P.O. Box 129188, Abu Dhabi, United Arab Emirates}

\author[0000-0003-2417-5975]{Glennys R Farrar}
\affiliation{Center for Cosmology and Particle Physics, Physics Department, New York University,
        New York, NY 10003}

\author{Sana Elgamal}
\affil{New York University Abu Dhabi,
    P.O. Box 129188, Abu Dhabi, United Arab Emirates}






\begin{abstract}

We have constructed a catalog of active galactic nuclei (AGNs) with $z < 0.13$, based on optical spectroscopy, from the parent sample of galaxies in the 6dF galaxy survey (Final Release of 6dFGS), a census of the Southern hemisphere. This work is an extension of our all sky AGN catalog \citet [ZCF, here after]{ZCF19}. The ZCF is based on 43,533 galaxies with $\rm K_s \leq$ 11.75 ($z \leq 0.09$) in
the 2MASS Redshift Survey (2MRS). The parent catalog of this work, the 6dF catalog, consists of 136,304 publicly available digital spectra for 125,071 galaxies with $\rm Dec \leq 0^\circ$ and $\rm K_s \leq 12.65$ (median $z = 0.053$). 
Our AGN catalog consists of 3109 broad line AGNs and 12,156 narrow line AGNs which satisfy the \citet{Kauffmann03} criteria, of which 3865 also satisfy the \citet{Kewley01} criteria. We also provide emission line widths, fluxes, flux errors, and signal-to-noise ratios of all the galaxies in our spectroscopic sample, allowing users to customize the selection criteria. 
In addition, we provide AGN likelihood for the rest of galaxies based on the availability and quality of their spectra. These likelihood values can be used for rigorous statistical analyses.

\end{abstract}




\section{Introduction} 
\label{sec:intro}

A complete census of active galactic nuclei (AGNs) is important for understanding the growth of supermassive black holes (SMBHs) and the co-evolution of SMBHs and their host galaxies via AGN feedback. 
While deep surveys have the highest number of AGNs, nearby, all-sky catalogs are important for 
understanding the low redshift universe. 
AGN catalogs can be used to search for and study AGN-related phenomena, such as circumnuclear water maser emission  \citep[e.g.,][]{Zhu11} and ultra-high energy cosmic rays (UHECRs) \citep[e.g.,][]{AugerSci, AugerCor}. The interpretation of statistical correlation studies between the arrival direction of UHECRs and the sky locations of AGNs depends on the completeness, homogeneity, and purity of the AGN catalogs used. 
Furthermore, nearby AGNs can be studied in greater detail and newly identified AGNs can serve as additional targets for high angular resolution studies, especially for powerful telescopes in the Southern hemisphere such as {ALMA  \citep{Kurz02}, MATISSE \citep{Lopez14}, and GRAVITY \citep{Gravityco}. 

Analysis of optical emission lines is one of the most robust methods of identifying AGNs, either via the presence of a broad Balmer line for Type 1 (unobscured) AGNs or via the ratios of the fluxes of the forbidden lines to those of Balmer lines for Type 2 (obscured) AGNs. Optical lines also provide measures of the AGN activity, since the [OIII] luminosity is a proxy for AGN bolometric luminosity, and of AGN obscuration, since the ratio of the H$\alpha$ to H$\beta$ line fluxes, called the Balmer decrement, increases with obscuration. 
Comparisons of the AGNs identified via different wavelengths is necessary to further  assess the strengths and weaknesses of each method as well as a better understanding of the physical mechanisms responsible for the differences \citep[e.g.,][]{Hickox18}.  While optical observations of AGNs are highly reliable, due to their rich line information, they often miss low luminosity and obscured AGNs. In comparison, infrared observations offer low obscuration bias but may miss star-formation dominated AGNs and dust poor AGNs. Similar to optical, X-ray observations are highly reliable and have low host contamination and low obscuration bias. However, X-ray observations can also miss low luminosity and heavily obscured AGNs. If the AGN has no host contamination, and sufficient radio AGN emission, then the radio wavelength band should be ideal for obscuration bias-free analysis. However, only a small fraction of AGNs are radio-loud.  AGNs identified with different methods should be combined to get a complete sample or as close as possible \citep[e.g.,][]{{Asmus2020}}. Our catalog can contribute to these studies, especially when compared with AGN catalogs from future X-ray telescopes, such as {\it eROSITA}{\footnote{\url{ http://www.mpe.mpg.de/eROSITA}}}, {\it Athena}{\footnote{\url{ http://www.the-athena-x-ray-observatory.eu/}}}, {\it Lynx}{\footnote{\url{ https://wwwastro.msfc.nasa.gov/lynx/}}}, and {\it XARM}{\footnote{\url{ https://heasarc.gsfc.nasa.gov/docs/xarm/}}}, and Far-IR-radio surveys such as ALMA{\footnote{\url{ http://www.almaobservatory.org/}}} and OST{\footnote{\url{https://asd.gsfc.nasa.gov/firs/}}} .

We have identified AGNs, using optical line ratios, from the Final Release of the 6dF Galaxy Survey \citep[6dFGS,][]{6dF04, 6dFGS}, a census of nearby galaxies in the Southern hemisphere with $\rm Dec \leq 0^\circ$ and $K_s \leq 12.65$. This work is an extension of the \citep[ZCF, here after]{ZCF19} all-sky, optical AGN catalog identified from the parent sample of 2MASS Redishift Survey \citep[2MRS, here after]{2MRS} galaxy catalog, which contained 6dF galaxies with $K_s \leq 11.75$. The Final Release of 6dFGS sample consists of 136,304 spectra (125,071 galaxies), more than ten times larger than the subset of 11,762 galaxies included in 2MRS.

This paper is organized as follows. Section~\ref{sec:sample} describes our spectroscopic sample. 
Section~\ref{sec:AGNID} describes the methods we used to isolate the emission lines and identify AGNs. Section~\ref{sec:catalog} describes our AGN catalog. We present the impact of spectral signal-to-noise on AGN detection rates and statistically correct for the resulting incompleteness and inhomogeneity of AGN detection rates in Section~\ref{sec:agnrates}. In Section~\ref{sec:conclusions} we summarize our catalog and findings.

\begin{figure}[h]
\begin{center}

\includegraphics[width=0.4\textwidth,angle=90]{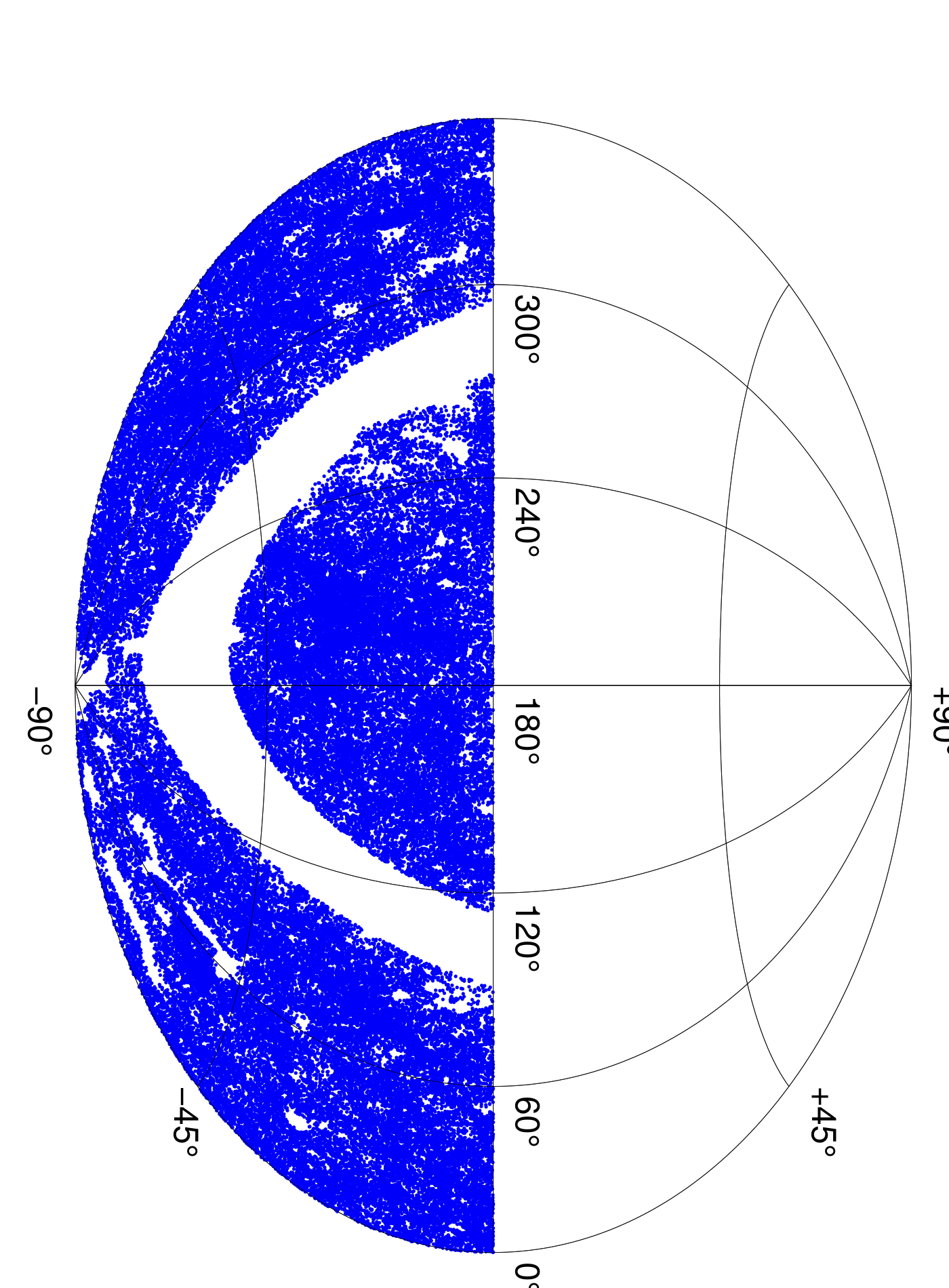}

\end{center}
\caption{The sky distribution of the galaxies in the Final Release of the 6dF Galaxy Survey.}
\label{fig:fullskycoord}
\end{figure}

\begin{figure}[htb]
\begin{center}
\includegraphics[scale=0.6,angle=0]{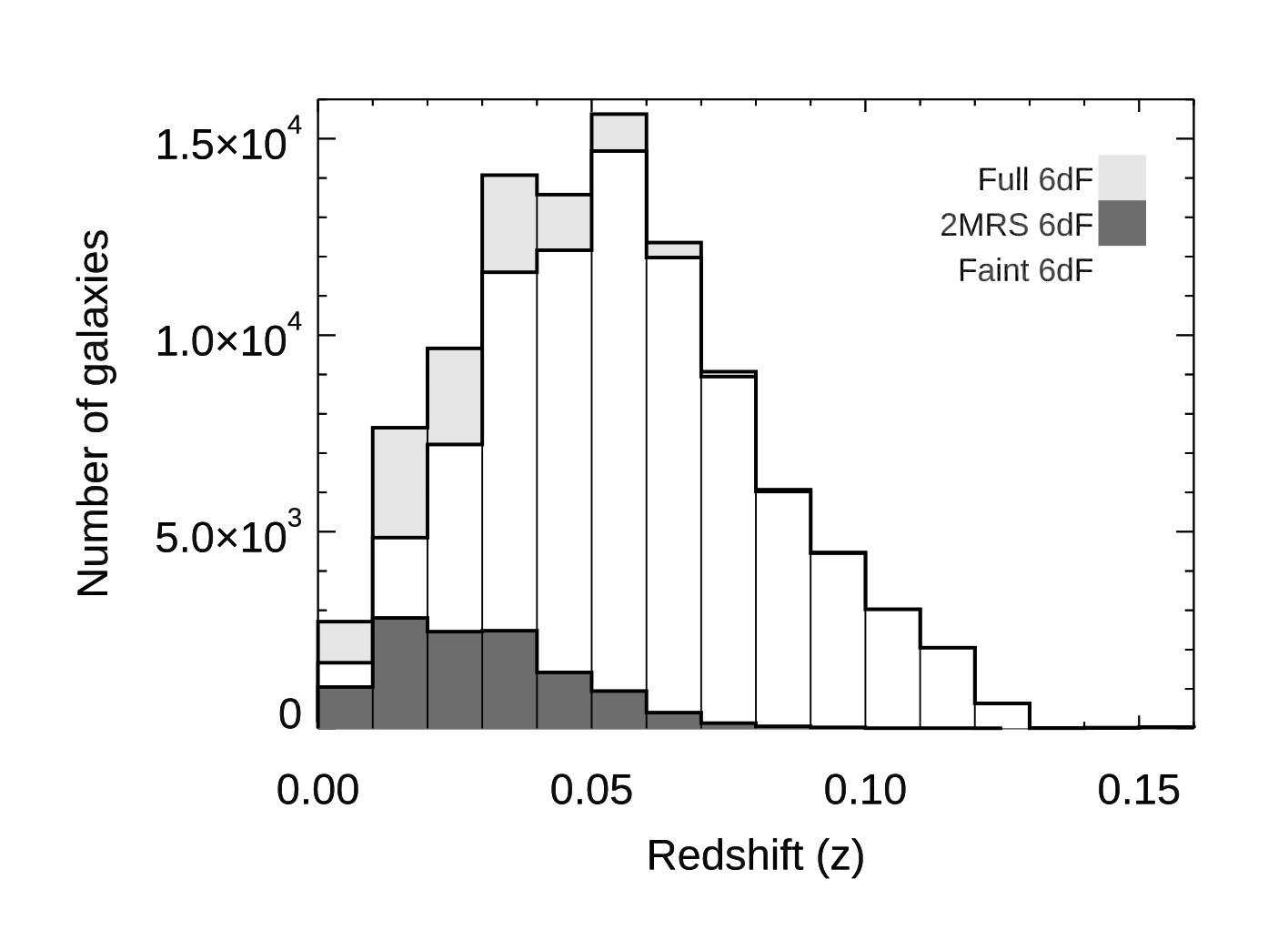}

\end{center}
\caption{Redshift distributions of the Full 6dF sample, 2MRS 6dF subsample (dark gray) and the Faint 6dF subsample.
To distinguish the different sample of the different set 6dF galaxies, we call the whole 6dF as ``Full-6dF'', 
the subset of 6dF data in 2MRS is ``2MRS-6dF'',
the 6dF galaxies that are not in the 2MRS and are fainter is ``Faint-6dF''.
Redshifts are adopted from \citet{6dF04, 6dFGS}.}
\label{fig:histredshiftcmp}
\end{figure}

\section{Spectroscopic Sample of Galaxies}
\label{sec:sample}

We take the Final Release of the 6dF Galaxy Survey \citep[6dFGS,][]{6dF04, 6dFGS} as the parent sample for constructing our Southern-sky optical AGN catalog.
The survey targeted galaxies in the Southern hemisphere ($\rm Dec \leq 0^\circ$) with  $K_s \leq 12.65$ (median $z = 0.053$) from the Two Micron All Sky Survey \citep[2MASS,][]{2MASS} catalog.
The 6dFGS catalog consists of 136,304 publicly available digital spectra of 125,071 galaxies taken with the Six-Degree Field (6dF) multi-fibre spectrograph, operated on the United Kingdom Schmidt Telescope (UKST). Figure~\ref{fig:fullskycoord} shows the sky distribution of the galaxies in the 6dFGS. 
 The Galactic latitude cut of 6dF data is $|b| > 10^{\circ}$, as described by \citet{6dF04}, 
and the same is used for this work.
The observations of 6dFGS were carried out at the UKST from 2001 to 2006 \citep{6dF04}. 
The majority of observations of the 6dFGS galaxies consist of two spectra, in V-band and R- band. Due to different gratings, data 
taken prior to 2002 October have V-band spanning 4000--5600 \AA\ and R-band spanning 5500--8400\AA; later 
observations ($\sim 80\%$) have V-band spanning 3900--5600 \AA\ and R-band spanning 5300--7500\AA.
V-band spectra have a resolution of 5--6 \AA\  in full wavelength at half maximum (FWHM), and R-band spectra have a resolution of 9--12 \AA\ FWHM.
 The data do not have absolute flux calibration.

In order to run the optical spectroscopic analysis for AGN candidates, we need spectra in both V-band, which covers the \Hb\  and $\rm [O\ III]$ lines,
and the R-band, which covers the \Ha\  and $\rm [N\ II]$\ lines. 
There are $\sim 3,460$ spectra with only one (either V or R) band available. 
Moreover, we find that there are $\sim$ 3,400 spectra with negative mean signal-to-noise (S/N) in either V- or R- band. We remove the spectra which only have one band available and the ones with negative mean S/N before we combine the V-and R- band data. 
To exclude the low S/N cases of absorption lines, we made a conservative cut at mean $\rm S/N \geq 3$ at both V- and R- band, where the  $\rm S/N$ is defined as the mean signal-to-noise over the full V-band or R-band. The data that passed the above criteria were then stitched to combine their V- and R- bands.  We note that there are some galaxies with repeated observation -- less than 10\% of the total number of galaxies.  A majority of these have two observations and a few have up to four observations.
In case of repeated observations for the same galaxies ($\sim 9,000$ galaxies), we keep the spectra with the highest S/N. After the above selection, we have 101,263 galaxies whose spectra passed the data quality control\footnote{The merged spectra would be made available as a tar file on request.}. 

\begin{figure}
\begin{center}
\includegraphics[scale=0.6,angle=0]{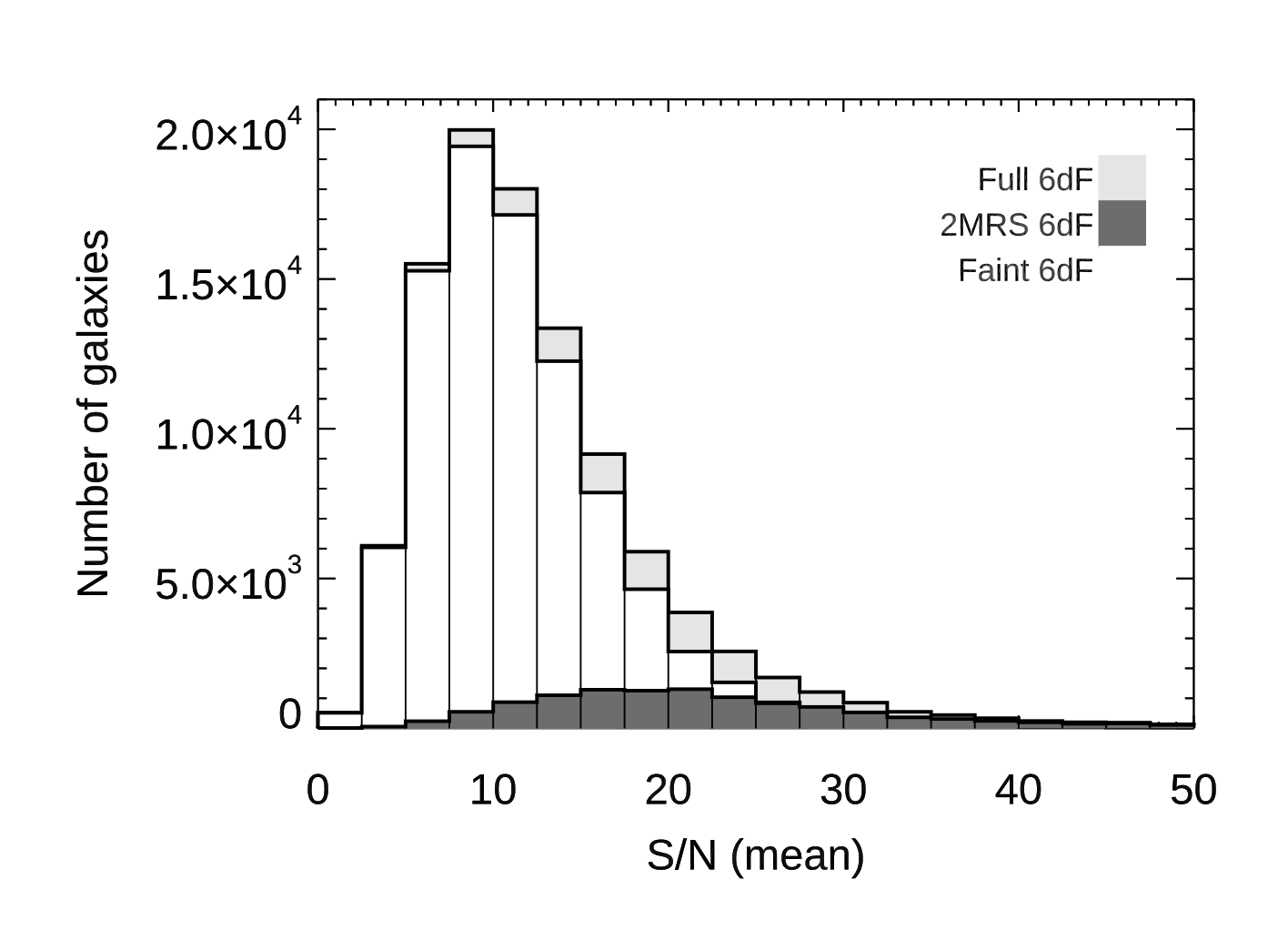}

\end{center}
\caption{Distribution of the mean signal-to-noise of the Full 6dF spectra (light gray), 2MRS 6dF spectra (dark gray), and Faint 6dF spectra (white).
To distinguish the different sample of the different set 6dF galaxies, we call the whole 6dF as ``Full-6dF'', 
the subset of 6dF data in 2MRS is ``2MRS-6dF'',
the 6dF galaxies that are not in the 2MRS and are fainter is ``Faint-6dF''.}
\label{fig:histmeansn}
\end{figure}

\begin{figure}[htb]
\begin{center}
\includegraphics[scale=0.5,angle=90]{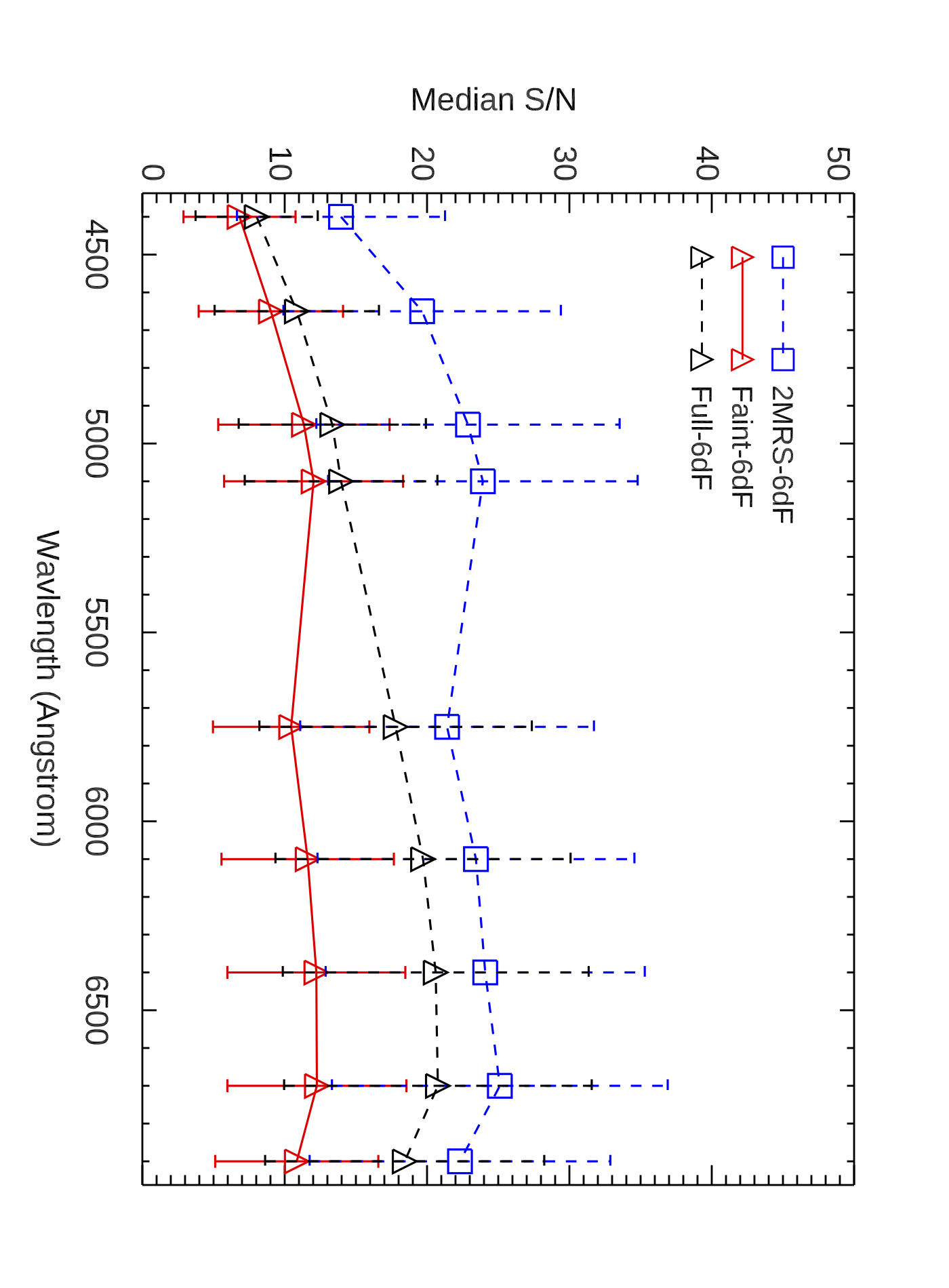}
\end{center}
\caption{Signal-to-noise ratio (S/N) of the subsamples vs. wavelength. 
To distinguish the different sample of the different set 6dF galaxies, we call the whole 6dF as ``Full-6dF'', 
the subset of 6dF data in 2MRS is ``2MRS-6dF'',
the 6dF galaxies that are not in the 2MRS and are fainter is ``Faint-6dF''.
We define the S/N in each wavelength bin as the data divided by the error provided in the error spectrum. The symbols are the mean and error bars are the standard deviation of the S/N distributions.}
\label{fig:sampleSNR}
\end{figure}

Note that a subset of the 6dF data (11,762 galaxies) were part of 2MRS \citep{2MRS} and the AGNs in that sample are part of our all sky optical AGN catalog \citep[ZCF]{ZCF19}. We do not repeat the data reduction and analysis of those here. The subset of 6dF data in 2MRS (hereafter 2MRS 6dF) and ZCF has a brighter magnitude cut at $K_s \leq 11.75$ to match the uniform sky distribution based on 2MRS survey. Therefore, the signal to noise of those spectra are higher than the Full 6dF sample, 
with a magnitude cut of  $K_s \leq 12.65$.
We show the redshift distribution in Figure~\ref{fig:histredshiftcmp}, where the subset of 6dF from 2MRS is highlighted in dark gray. Redshifts ($z$) are from \citet{6dF04, 6dFGS}. Due to the $K_s$ requirement of 2MRS compared to 6dF,
the objects of 2MRS 6dF are, in general, brighter than the rest of 6dF objects. 
We therefore separate the 6dF sample into 2MRS 6dF and Faint 6dF subsamples.
We show the mean S/N distribution over all wavelengths in Figure~\ref{fig:histmeansn}. The 2MRS 6dF subsample has an overall higher S/N, with the mean S/N peaked at 20, while the Full 6dF has significantly lower S/N, peaked at 10. 
Since the S/N values are wavelength dependent \citep[e.g.,][]{ZCF19}, we show the mean and standard deviations of the S/N for the 6dF sample at various  (continuum) wavelengths in Figure~\ref{fig:sampleSNR}, where subsamples from 2MRS and the Faint 6dF are presented to show the S/N differences. 

\begin{figure}
\centering
\includegraphics[scale=0.8,angle=0]{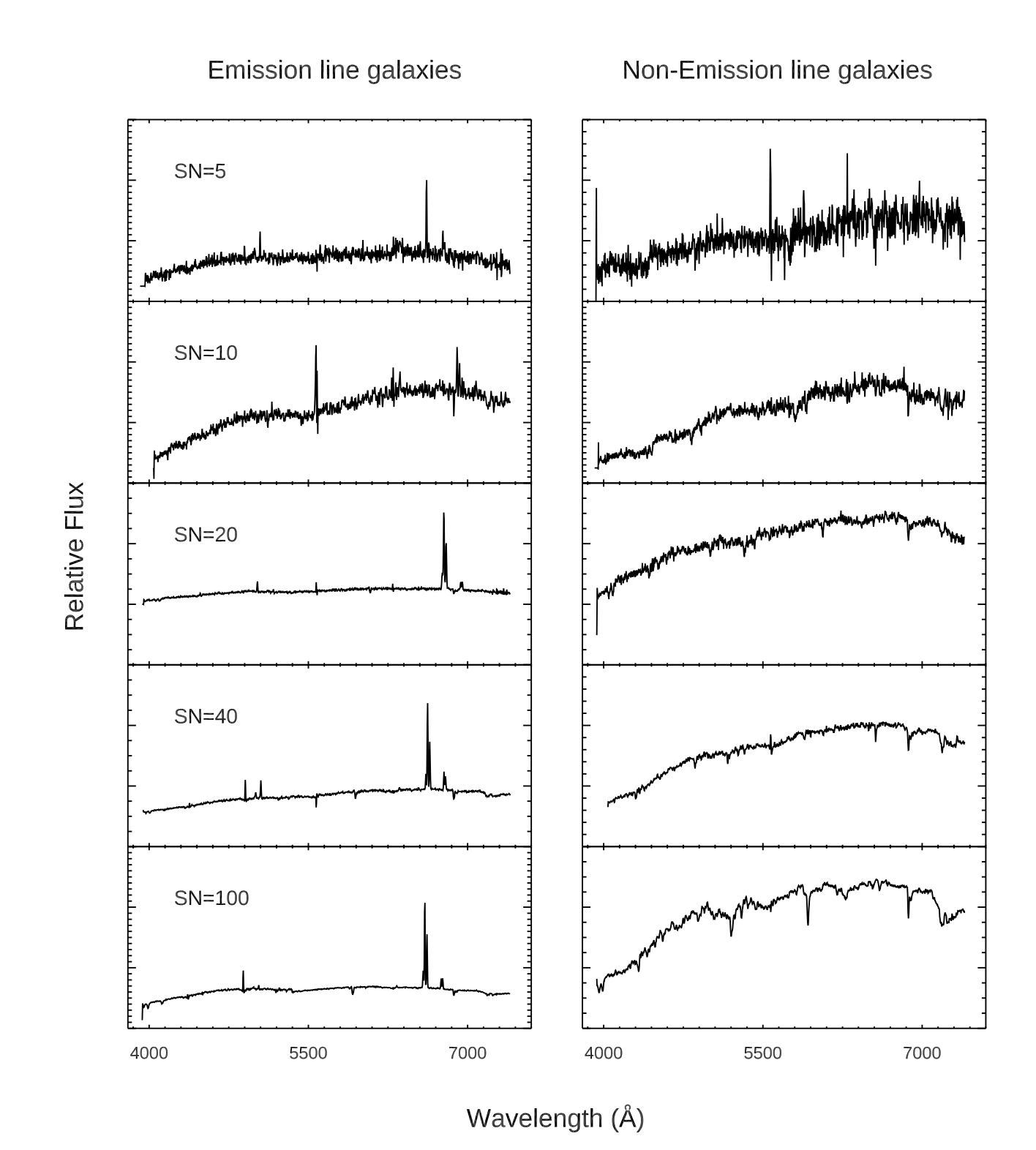}
\caption{
Sample spectra of 6dF data. Left column shows spectra of emission line galaxies, right column shows spectra of non-emission line galaxies. Mean signal-to-noise increases from top to bottom. }
\label{fig:sample_sp}
\end{figure}

We use the wavelength range covered by both V- and R- bands (5500--5600 \AA\ or 5300--5600 \AA) to merge the 6dF data that satisfy the  S/N criteria into a single 
spectrum for each galaxy. Data are evenly resampled to log space to reduce the possible pixel shifts between different bands
since there are some small wavelength shifts for the same spectral features in between the two arms due to the wavelength calibration procedure. We then use the arm with higher resolution, namely the V-band, as a template to fit for the overlapped wavelength ranges in the R-band using the Penalized Pixel-Fitting  \citep[pPXF,][]{Cappellari04} method to determine the number of pixels to be shifted and to merge the bands. Note that this is an automatic procedure, which stitches most of the data successfully. There are cases where the automatic stitching process fails due to poor data quality in one of the arms, or a large number of pixels 
in the overlap region whose values are zeros, especially for pixels at the beginning of the R-band and end of the V-band.
 Examples of arm-merged 6dF spectra with different continuum signal-to-noise ratios are shown in Figure~\ref{fig:sample_sp}.
We identified $\sim$ 7000 spectra which fail automatic stitching by visual inspection. For the these cases, we had to merge the data manually, i.e., scaling the arms and shifting the pixels until no obvious discontinuity appeared between the two bands.

We note that the spectra are not corrected for the Earth's atmospheric absorption, i.e., the telluric absorption. The B-band around 6860--6890 \AA\ overlaps with the optical AGN identification emission lines, $\rm [N\ II]$ -\Ha\  complex, and can cause inaccurate line fluxes. This typically happens to the objects in the redshift range  0.0407 $< z <$ 0.05111. In those cases, we exclude these galaxies ($\sim 14 \%$) from our sample. However, when a clear broad \Ha\ component is present, we still keep the galaxy as a potential AGN candidate although the fluxes of the broad  \Ha\  in these galaxies cannot be measured accurately due to the telluric contamination. In addition, when the galaxies are at higher redshifts $z \geq 0.1428 $, 
the \Ha\ line may shift out of the wavelength range, and we cannot assess whether they are AGNs or not.

\section{Spectral Fitting and AGN Identification}
\label{sec:AGNID}

The spectral fitting is performed on the bands-merged spectra in order to 
isolate the emission lines used in optical AGN identification.
The spectral fitting procedure and validations are 
described in detail by \citet{ZCF19}. We briefly summarize it here. 

\subsection{Subtraction of Host Galaxy Contribution to the Spectra}

We subtract the stellar absorption and continuum emission from the galaxy by modeling them using full spectrum fitting with a stellar population model. As various of stellar population models are available, using different template models may result in systematic differences in AGN identification \citep{CZF}.
The \citet{MILES} stellar population models are based on the empirical stellar library MILES that populates the current largest stellar parameter space.
Therefore, we use the \citet{MILES} stellar population models, which cover the wavelength range from 3525 to 7500\AA. The model spectra have a resolution of full-wavelength-at-half-maximum (FHWM) 2.5 \AA. We broaden the model spectra to match the observed spectral resolution of 6dF data. Each 6dF spectrum is fit as a linear combination of the single stellar population (SSP) models using the pPXF\citep{Cappellari04} program. 
 We use the redshift from \citet{6dF04, 6dFGS} as initial guesses in our SSP fits and refit for the recessional velocity of each galaxy. The results we derived are consistent with the velocity from \citet{6dF04, 6dFGS}, 
The mean value of the velocity difference distribution is 26\kms; the standard deviation of the velocity difference distribution is 79\kms, roughly half the spectral resolution (150\kms) of 6dF spectra. 
We applied same SSP $\chi^2$ cut at 6.05 to keep 99\% of the successful fits as in ZCF \citep{ZCF19}.

Due to the data quality or extremely strong emission features, about $\sim$19 percent of galaxies failed in the template fitting procedure. 
A failed fit either has no template or yields an over-broadened best-fit template (with Gaussian kernel broadening larger than 1,000 \kms).
In these cases, we fit for the continuum in the wavelengths on either side of the emission lines of interest and interpolate; no \Ha\  and \Hb\  stellar absorption contributions are subtracted. If the failure is due to the fact that there is no significant absorption along the wavelength then no bias is introduced. If the failure is due to low quality of the spectra, i.e., low signal-to-noise, the emissions of \Ha\  and  \Hb\  are underestimated, hence the line ratios are overestimated. This may lead to a systematically higher number of galaxies identified as AGNs \citep{ZCF19}. We mark the AGNs that are identified after this ``local fitting'' procedure in the catalog. Readers should treat these objects with caution.

\subsection{AGN Identification}
\label{sec:lineratios}

After subtracting the continuum and absorption contributions from the host galaxy, we identify AGNs from the emission lines in the spectrum. Type 1 (broad line) AGNs are characterized by the presence of broadened hydrogen Balmer lines. As explained by the unified model 
\citep[e.g.,][]{Antonucci93, Urry95} this is the case when an AGN is unobscured, i.e. viewed face-on, and the Doppler broadened emission of the rapidly orbiting gas close to the supermassive black hole can be seen in the spectrum. If an AGN is seen closer to edge-on, the broad line regions are obscured by intervening gas and dust, leaving only narrow emission lines in the spectrum. In this case, the Type 2 (narrow line) AGNs can be identified from the flux ratios of the forbidden ([NII] and [OIII]) lines to the Balmer (\Ha\  and \Hb\ ) lines.

We fit Gaussians to the emission lines in the spectrum and calculate the full width at half maximum (FWHM) and flux of each line.
For the $\rm [N\ II]$ -\Ha\ complex where the lines are blended together, three or four Gaussians are used to fit for the emission components simultaneously. 
The corresponding Gaussian components are used to calculate the fluxes and FWHM of $\rm [N\ II]$ and \Ha\ lines. 
The details of the fitting procedure are described in Section 3.3 and shown in Figure 6 of \citet{ZCF19}. 
Line fluxes are calculated under the Gaussian profiles in the wavelength bins within 3$\sigma$ of the fitted peak, where $\sigma$ is the fitted width of the Gaussian. Flux errors are calculated by adding in quadrature the value of each bin within the same wavelength range (i.e. 3$\sigma$) in the error spectrum. 

Our method has been validated by comparing it with the results from SDSS Data Release 8 \citep[DR8]{SDSS8}. The detailed description can be found in \citet{ZCF19} and we briefly summarize the main procedures and results here. We first compared the fluxes and flux errors we derived for the SDSS galaxies in 2MRS with those from DR8 and found good agreement. We then compared our measurements for the galaxies which have spectra from both SDSS and 6dF. Since 6dF spectra do not have absolute flux calibration, we cannot compare the fluxes. However, we find that the flux ratios are consistent between SDSS and 6dF spectra for the same galaxies. Since 6dF spectra have lower signal-to-noise on average, requiring $\rm S/N \geq 2.0$ for emission lines in 6dF spectra gives roughly the same AGN identification rates as requiring $\rm S/N \geq 3.0$ for emission lines in SDSS spectra.

Because redder diagnostic lines such as $\rm [S\ II]\ \lambda\lambda$ 6717, 6731 lines may shift out of the available spectral wavelength at high redshift ($z \geq 0.114$), for completeness, we only use the bluer optical emission lines, i.e.,  broad and narrow \Ha\ , narrow \Hb\ ,  $\rm [O\ III]\ \lambda5007$, and $\rm [N\ II]\ \lambda6584$, for AGN identification.
We use the ratio between flux and flux error, $\rm (S/N)_{line}$, to evaluate the significance of the emission line features. 
Following the detailed discussion in \citet{ZCF19}, where common galaxies with high S/N are available from SDSS,  $\rm (S/N)_{line}\geq 2.0$ is used to further investigate the AGN candidates. An emission line galaxy is defined as one having all four lines with $\rm (S/N)_{line}\geq 2.0$.
When broad \Ha\ has $\rm (S/N)_{line} < 2.0$ while the other narrow lines all have $\rm (S/N)_{line}\geq 2.0$, a triple Gaussian is used to refit the $\rm [N\ II]$ -\Ha\ complex to derive the corresponding line fluxes and flux errors. Galaxies with significant ($\rm (S/N)_{line}\geq 2.0$) broad \Ha\ components and having wide FWHM $\geq$ 1000 \kms \citep[e.g.,][]{Eun17, Ho97, Schneider10, Stern12, VandenBerk06} are classified as Type 1 AGNs. As mentioned above, the $\rm [N\ II]$ -\Ha\ complex is contaminated by the telluric absorptions at certain redshifts. The telluric contaminated galaxies can be potential Type 1 AGNs. When such a galaxy shows a wide enough and prominent (usually through visual identification) broad \Ha\, we still identify it as a Type 1 AGN, although its broad \Ha\ flux is underestimated.
Normally \Hb\ lines are weaker than \Ha\ lines, therefore we do not require broad components of \Hb\ lines as the indication of Type 1 AGNs. However, there are potential Type 1 AGNs with broad \Hb\ components but weaker broad \Ha\ as well. Those (about a few hundred) are semi-manually identified by checking their FWHM, $\rm (S/N)_{line}$, and visual inspection. As mentioned in Section~\ref{sec:sample}, no flux-calibration was performed on the data, therefore, absolute flux values are not available in this work. 

When no significant broad \Ha\ or \Hb\ present, we use the line flux ratio as first proposed by \citet{BPT} known as the BPT diagram to distinguish the Type 2 (``narrow line") AGN from the star-forming galaxies whose spectra also show emission features from $\rm [O\ III]$, $\rm [N\ II]$, \Ha, and \Hb. Line ratios \rm{[NII]}/\Ha\ and \rm{[OIII]}/\Hb\ are used in the BPT diagram to identify the possible AGN candidates. 
We adopt the two demarcation lines, which are usually used to separate AGNs and s\Hb\ tar forming galaxies, namely,

\begin{alignat}{2}
&\log(\rm{[OIII]}/\Hb) && > 0.61/(\log(\rm{[NII]}/\Ha) - 0.47) + 1.19 \\
&\log(\rm{[OIII]}/\Hb) && > 0.61/(\log(\rm{[NII]}/\Ha) - 0.05) + 1.3. 
\end{alignat}

The first is based on theoretical modeling of maximal line ratios possible in star formation \citep{Kewley01}, and the second is based on empirical studies of SDSS galaxies \citep{Kauffmann03}. Galaxies with line ratios falling above the \citet{Kewley01} line have forbidden line emission dominated by the AGN. Galaxies between the \citet{Kewley01} and \citet{Kauffmann03} lines are likely to be Type 2 AGN but can 
have contribution from both AGN and star formation galaxies. These galaxies are also referred as ``composite" galaxies. We use the \citet{Kauffmann03} line to identify AGNs for our catalog. If an AGN is also above the \citet{Kewley01} line, we mark it as such.

\begin{figure}
\begin{center}
\includegraphics[width=0.35\textwidth, angle=90]{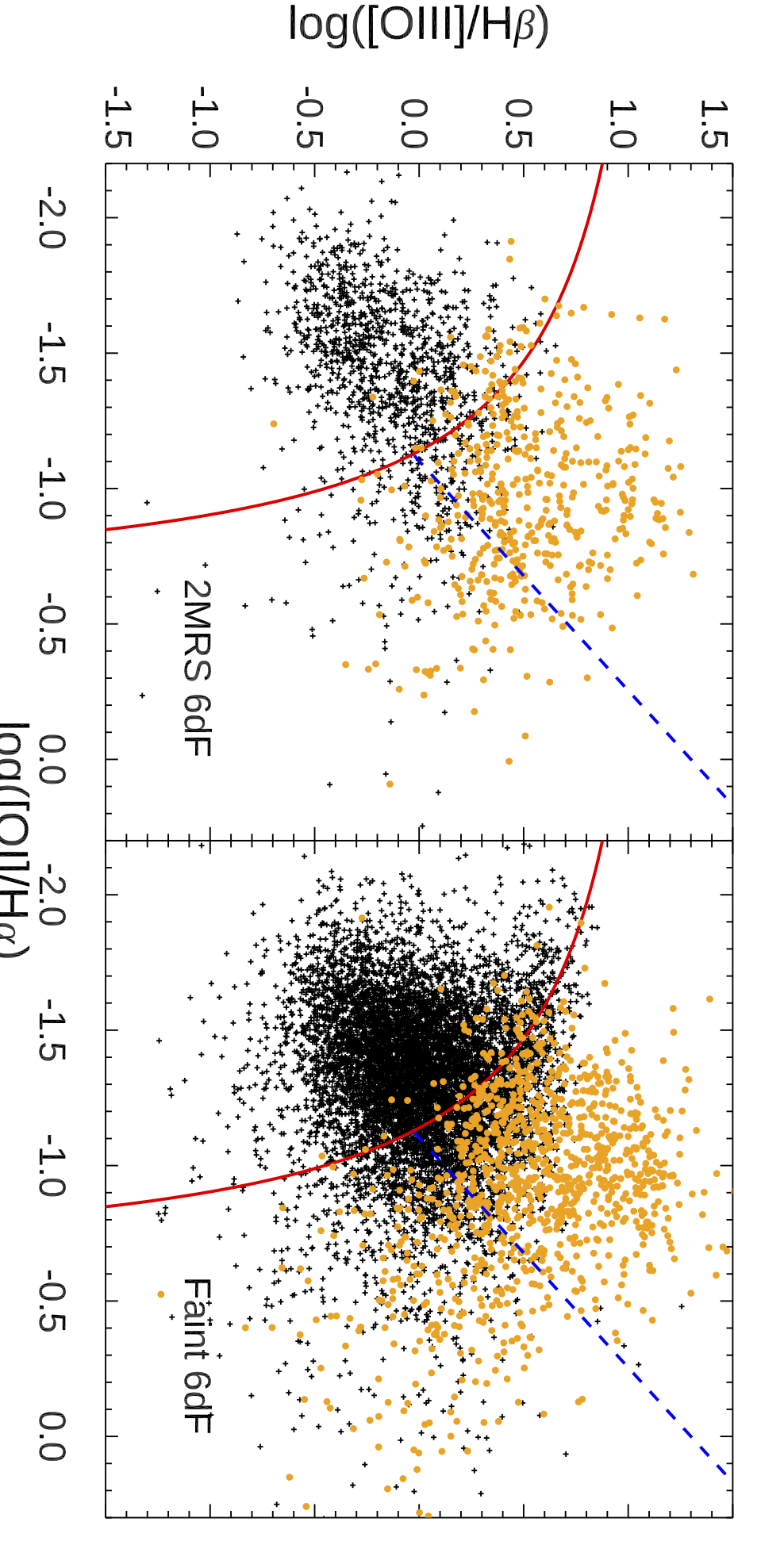}
\caption{The Type 2 AGNs in our catalog requiring that $\rm [O_I]$ line has $\rm S/N \geq 2.0$ in addition to the four main AGN identification lines, i.e., \Ha\ , \Hb\ ,  $\rm [O\ III]\ \lambda5007$, and $\rm [N\ II]\ \lambda6584$. The orange dots are AGNs and black dots are non AGNs according to the  \rm{[NII]}/\Ha\ criteria from \citet{Kewley01}. The red line shows the $\rm [O_I]$ AGN demarcation line and the blue dashed line shows the Sy2 vs. LINERs line from \citet{Kewley06}.
}
\end{center}
\label{fig:AGN_liner2_2p}
\end{figure}

\begin{figure}
\begin{center}
\includegraphics[width=0.35\textwidth, angle=90]{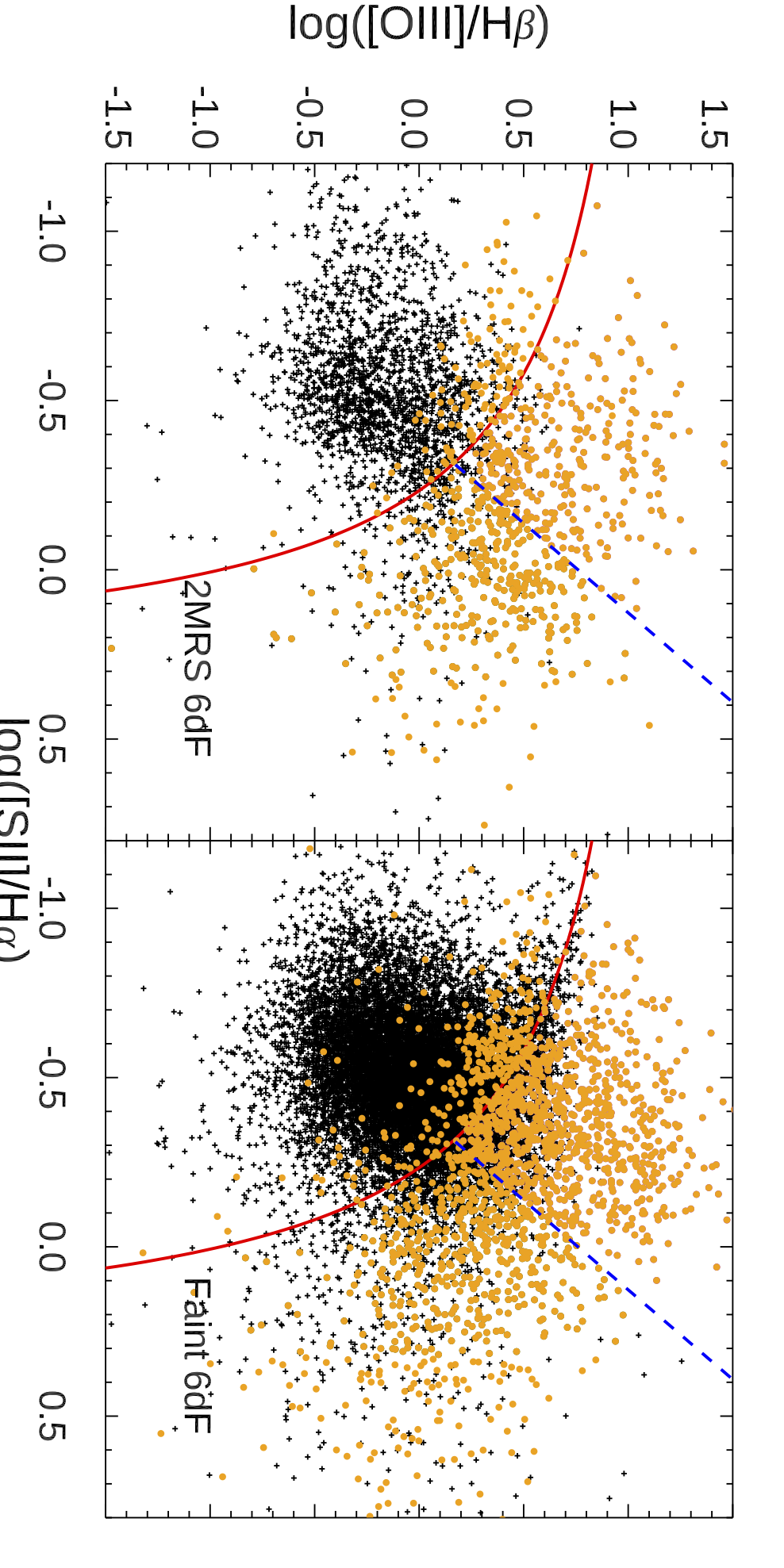}
\caption{The Type 2 AGNs in our catalog requiring that $\rm [S_{II}]$ line has $\rm S/N \geq 2.0$ in addition to the four main AGN identification lines, i.e., \Ha\ , \Hb\ ,  $\rm [O\ III]\ \lambda5007$, and $\rm [N\ II]\ \lambda6584$. The orange dots are AGNs and black dots are non AGNs according to the  \rm{[NII]}/\Ha\ criteria from \citet{Kewley01}. The red line shows the $\rm [S_{II}]$ AGN demarcation line and the blue dashed line shows the Sy2 vs. LINERs line from \citet{Kewley06}.} 
\end{center}
\label{fig:AGN_liner2_s2}
\end{figure}

 As there are multiple AGN identification line ratios, such as $\rm{[SII]}/\Ha$\ and $\rm{[OI]}/\Ha$ \citep{Kewley06} besides the main line ratio of $\rm{[NII]}/\Ha$, we explore the AGN identification consistency with different criteria for those galaxies whose $\rm{[SII]}$ and/or $\rm{[OI]}$ emission lines also have $\rm S/N \geq 2.0$. We show the results in Figures~\ref{fig:AGN_liner2_2p} and ~\ref{fig:AGN_liner2_s2}. In both plots, AGN identified from $\rm{[NII]}/\Ha$ are marked as orange dots. The AGN identification using both $\rm{[SII]}/\Ha$ and $\rm{[OI]}/\Ha$ are in general agreement with the results of $\rm{[NII]}/\Ha$. As expected, some deviations occur from using different line ratios, with some galaxies shift from being identified as AGNs to being identified as non-AGNs or vice versa. In general, only a small of fraction galaxies that were identified as AGNs by  $\rm{[NII]}/\Ha$ move into the non-AGN region when using $\rm{[SII]}/\Ha$\ and $\rm{[OI]}/\Ha$ line ratios. A larger fraction of galaxies which are not AGNs by $\rm{[NII]}/\Ha$ criteria are identified as AGNs in the $\rm{[SII]}/\Ha$\ and $\rm{[OI]}/\Ha$ BPT diagrams. This indicates that $\rm{[NII]}/\Ha$ provides a purer sample of AGNs. In addition, $\rm{[NII]}$ lines are stronger and have a lower wavelengths, i.e., falls within the 6dF wavelength ranges at higher redshifts than the $\rm{[SII]}$ and $\rm{[OI]}$ lines. Therefore, we use $\rm{[NII]}/\Ha$ as the AGN identification line ratio for this catalog.

\section{The Catalog}
\label{sec:catalog}

Our AGN catalog consists of 3116 Type 1 AGNs, and 12,162 Type 2 AGNs which satisfy the \citet{Kauffmann03} criteria. A subsample of 3866 Type 2 AGNs also satisfy the \citet{Kewley01} criteria. 
We summarize the AGN numbers and fractions of each subsample in Table~\ref{tab:AGN}. $N_{sample}$ is defined as the total number of valid spectra of each subsample, which includes the spectra that are free from telluric contaminations and some strong Type 1 AGN spectra affected by the telluric absorption. 
As mentioned above, we made a cut on S/N $\geq 3$ in V-band for Type 2 AGNs and S/N $\geq 3$ in R-band for Type 1 AGNs. 
A few Type 1 AGNs were identified by visual inspection whose R-band spectra are with S/N $\leq 3$ but strong broad emission lines.
We mark the objects in common with \citet{ZCF19} in this catalog by adopting their 2MRS name, i.e., name of TMID. Galaxies without TMID values are identified AGN from the Faint-6dF sample.
The sky distribution of our AGN catalog is shown in Figure~\ref{fig:AGNskyplot}. We show the BPT diagram for narrow emission line galaxies in Figure~\ref{fig:AGNBPT}, where line flux signal-to-noise cut  $\rm (S/N)_{line}\geq 2.0$ is applied to all the four diagnostic lines. 
In Figure~\ref{fig:AGNz_vs_sn} we show the mean S/N of the AGN sample across the V-band and R-band as a function of redshift. Data quality quickly decreases with redshift, from a mean S/N of $\sim30$ at $z = 0.0$ to S/N of $\sim 13$ at $z = 0.06$.  AGNs with higher redshift $z \geq 0.06$ have mean S/N of $\sim10.$ 

We use the \citet{Kewley06} criteria to further delineate Seyferts and low-ionization nuclear emission line regions (LINERs). Note that $\sim 2\%$ of the AGN sample are missing their {\rm [SII]} lines at higher redshift, the discussion about LINERs is limited to lower redshifts $z \leq 0.114$ . 
We show the Seyferts and LINERs of 2MRS and Faint-6dF samples in Figure~\ref{fig:AGN_liner_2p}. The LINER to Seyfert ratio of 2MRS-6dF is higher than that from the Faint-6dF subsample. This is likely because LINERs have weaker emission lines and they are preferentially lost in spectra with lower S/N \citep{ZCF19}.  
 As expected, the LINER to Seyfert ratio of 2MRS dropped from 1.01 to 0.56 in the Faint 6dF sample which has much lower S/N.  
Note that we use the \rm{[OIII]}/\Hb\  vs. \rm{[NII]}/\Ha\  BPT diagram to identify AGNs, so some emission line galaxies which do not qualify as AGNs in the main diagram appear in the AGN region in the \rm{[OIII]}/\Hb\  vs. \rm{[SII]}/\Ha\ diagram.

\begin{figure}[htb]
\begin{center}
\includegraphics[scale=0.4, angle=90]{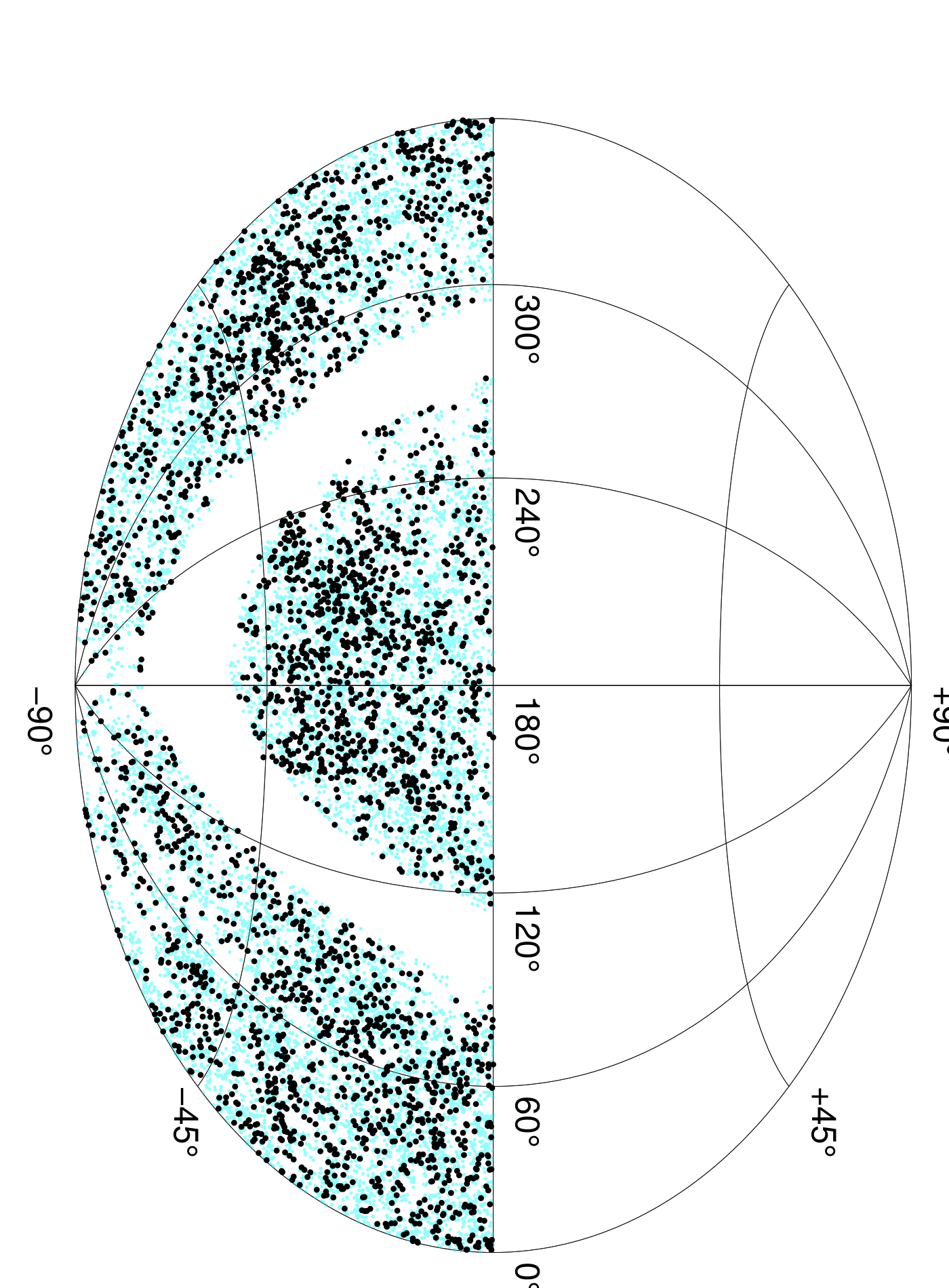}
\end{center}
\caption{The sky distribution of the AGN catalog from 6dF Type 1 (broad line) AGNs and Type 2 (narrow line) AGNs satisfying the \citet{Kauffmann03} criteria. The Type 1 AGNs are marked as black dots. }
\label{fig:AGNskyplot}
\end{figure}

\begin{figure}[htb]
 \begin{center}
\includegraphics[scale=0.6]{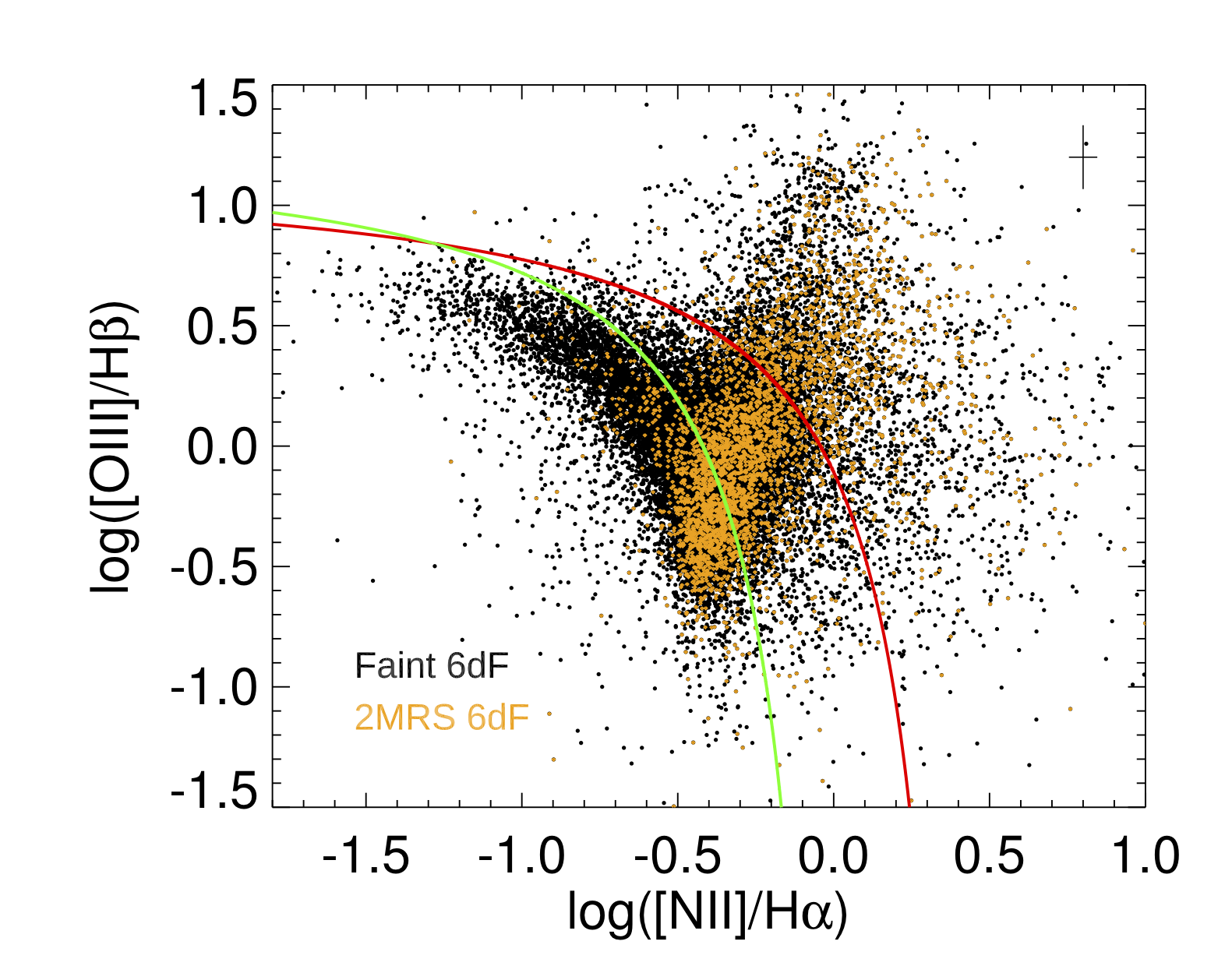} 
\end{center}
\caption{The BPT diagrams for the emission line galaxies ($\rm (S/N)_{line} \geq 2.0$ 
for all four lines). 6dF galaxies in 2MRS are shown as orange dots for comparison. The green and red lines denote the \citet{Kauffmann03} and \citet{Kewley01} criteria for narrow line AGNs, respectively.}
\label{fig:AGNBPT}
\end{figure}

\begin{figure}[thb]
\centering
\includegraphics[width=0.5\textwidth, angle=90]{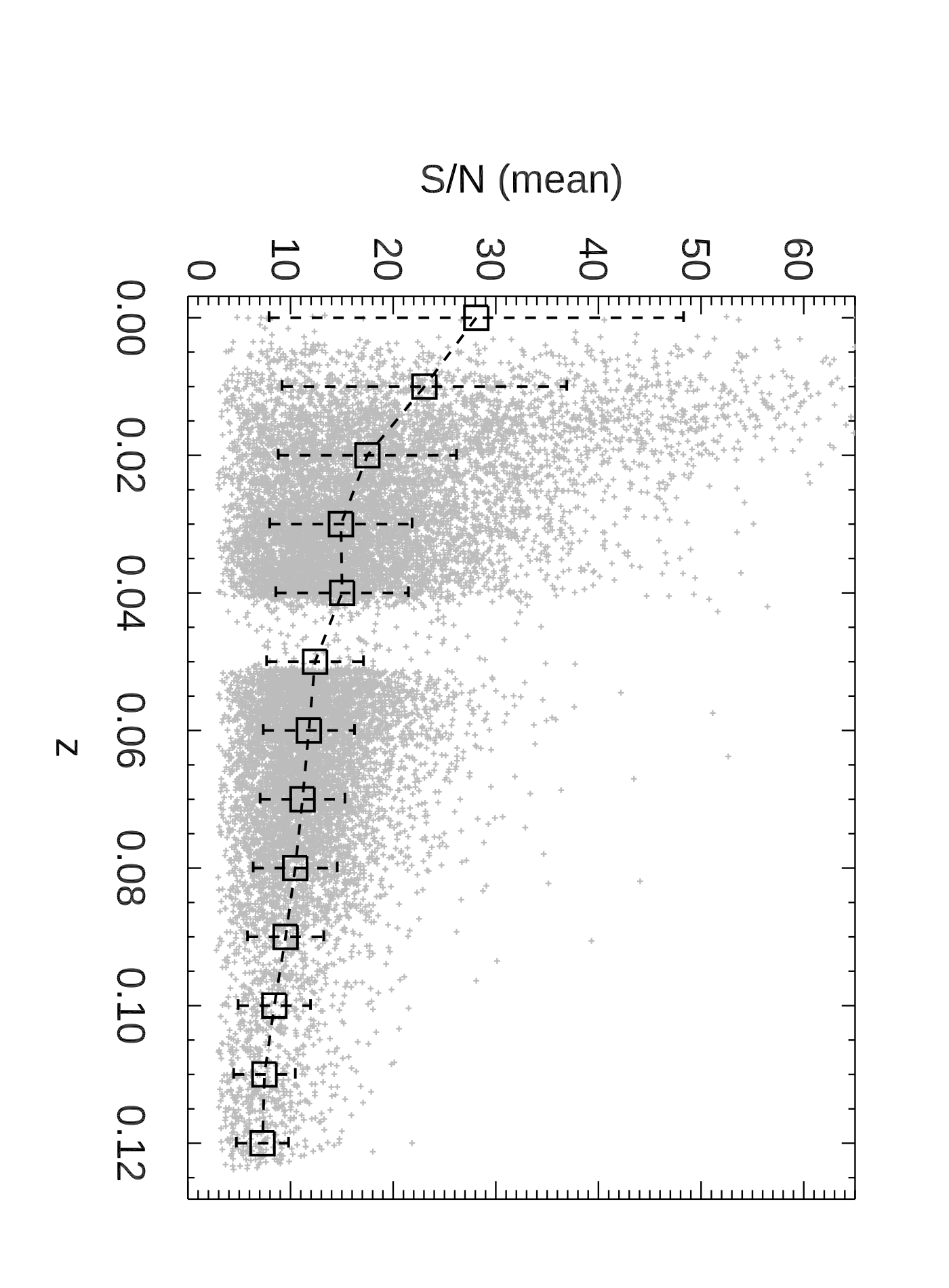}
\caption{Mean S/N of the AGN sample as a function of redshift. The mean value in each redshift bin is shown as a square. Error bars are the standard deviations of each bin.}
\label{fig:AGNz_vs_sn}
\end{figure}

\begin{figure}[thb]
\centering
\includegraphics[scale=0.5, angle=90]{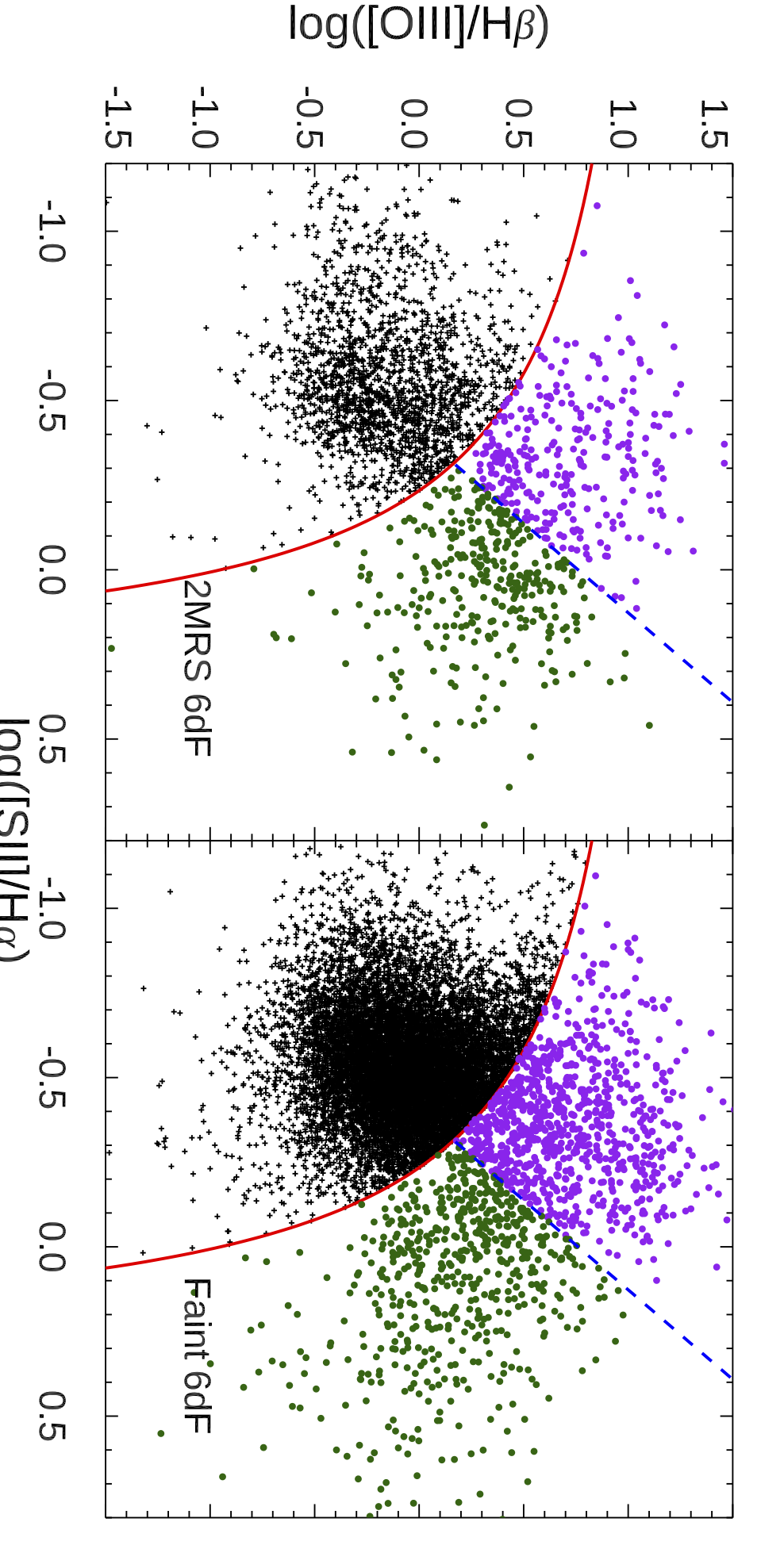}

\caption{The Type 2 AGNs in our catalog, separated into Seyferts (purple) and LINERs (green) using the \citet{Kewley06} criteria. There are more LINERs than Seyferts in the 2MRS 6dF subsample ($\rm K_s \leq 11.75$ ), and more Seyferts than LINERS in the Faint 6dF subsample ($\rm K_s \leq 12.65$) due to the worse data quality.
LINERs have weaker emission lines, so they are preferentially lost when spectral signal-to-noise decreases, i.e., the Faint 6dF galaxies. Note that we use the \rm{[OIII]}/\Hb\  vs. \rm{[NII]}/\Ha\  BPT diagram to identify AGNs, so some emission line galaxies which do not qualify as AGNs in the main diagram appear in the AGN region in the \rm{[OIII]}/\Hb\  vs. \rm{[SII]}/\Ha\ diagram.} 
\label{fig:AGN_liner_2p}   
\end{figure}

\begin{table}
 \caption{The AGN Catalog (Illustrative Extract)} 
\label{tab:catalog}    
 \begin{tabular}{l r r l l r r l r} 
\hline\hline
  \multicolumn{1}{c}{TARGETNAME} &
  \multicolumn{1}{c}{RA \ (deg)} &
  \multicolumn{1}{c}{DEC\ (deg)} &
  \multicolumn{1}{c}{$z$} &
  \multicolumn{1}{c}{Type} &
  \multicolumn{1}{c}{SN$_V$} &
  \multicolumn{1}{c}{SN$_R$} &
  \multicolumn{1}{c}{TMID} &
  \multicolumn{1}{c}{SSP $\chi^2$} \\
   &   &   &   &   &   &   &   &  \\ 
\hline
  g0000113\_050932 & 0.04709 & -5.15876 & 0.019057 & K03 & 12.6 & 9.6 & 00001131-0509313 & 1.26\\
  g0000182\_472923 & 0.07596 & -47.48958 & 0.027322 & K01 & 23.6 & 23.6 & 00001825-4729226 & 0.99\\
  g0000261\_490431 & 0.10858 & -49.07539 & 0.067518 & K01, Sy2 & 12.8 & 12.0 & & 1.23 \\
  g0000356\_014547 & 0.14849 & -1.76318 & 0.024424 & K03 & 19.9 & 17.9 & 00003564-0145472 & 1.56\\
  g0001138\_440043 & 0.30740 & -44.01183 & 0.038964 & K03 & 17.1 & 15.5 & 00011378-4400426 & 0.92\\
  g0001361\_144455 & 0.40020 & -14.74867 & 0.037679 & K01, LINER & 21.5 & 28.8 & 00013605-1444548 & 1.14\\
  g0001558\_273738 & 0.48258 & -27.62723 & 0.028330 & K01 & 14.1 & 17.8 & 00015583-2737382 & 2.07\\
  g0002039\_332802 & 0.51617 & -33.46728 & 0.028937 & K03 & 33.4 & 32.6 & 00020386-3328023 & 1.37\\
  g0002348\_034239 & 0.64505 & -3.71072 & 0.021498 & Sy1 & 21.4 & 16.6 & 00023480-0342386 & 1.84\\
  g0054323\_404258 & 13.63458 & -40.71603 & 0.024142 & K03 & 26.7 & 27.3 & 00543231-4042578 & 0.62\\
  g1211143\_393327 & 182.80946 & -39.55747 & 0.061064 & Sy1 & 17.3 & 18.8 & 12111425-3933268 & 0.77\\
  g1606162\_291654 & 241.56742 & -29.28167 & 0.015668 & K03 & 68.1 & 71.9 &  & 4.72\\
  g1901287\_243406 & 285.36954 & -24.56833 & 0.029157 & Sy1 & 27.1 & 28.5 & 19012871-2434060 & 1.41\\
  g1955064\_392224 & 298.77538 & -39.37258 & 0.019925 & K01 & 52.7 & 52.3 &  & 4.06\\
  g2124243\_234134 & 321.10283 & -23.69403 & 0.000307 & Sy1 & 5.1 & 10.9 &  & 0.99\\
  g2241589\_373443 & 340.49258 & -37.57853 & 0.028797 & Sy1 & 23.3 & 24.5 & 22415822-3734426 & 0.79\\
  g2312198\_274019 & 348.08262 & -27.67189 & 0.004918 & K01 & 3.8 & 3.7 &  & 0.78\\
  g2347461\_372925\# & 356.94221 & -37.49031 & 0.043774 & Sy1 & 22.9 & 25.9 &  & 0.72\\
  g1148026\_184952* & 177.01100 & -18.83114 & 0.033478   &  Sy1 & 3.8 & 3.9 &  & 3.19 \\
 g1526535\_310839t & 231.72292 &  -31.14408 &  0.047590 &  Sy1& 17.1 & 19.7 &15265348-3108388 & 1.07\\
\hline\end{tabular}
 \begin{tablenotes}
      \small
      \item TARGETNAME: 6dF ID, a \# indicates that ``local fitting" was used to estimate the galaxy contribution; 
      a * indicates Type 1 AGN identified from broad \Hb; a $``t''$ indicates Type 1 AGN may be contaminated by telluric absorption in the $\rm [N\ II]$ -\Ha\ complex;
      RA: right ascension; DEC: declination; $z$: redshift;  Type: AGN type, Sy1=Type 1, K01=Type 2 with Kewley et al. (2001) criteria, K03 = Type 2 with Kauffmann et al. (2003) criteria. We use $\rm [S_{II}]$ to classify the LINERs and Sy2 by adopting \citet{Kewley06} criteria. When $\rm [S_{II}]$ falls within the 6dF wavelength range, a K01 galaxy that satisfies either the LINERs or Sy2 criteria is marked accordingly in the column of `Type'; TMID: Two MASS ID, indicate common AGNs in ZCF \citep{ZCF19}; SSP $\chi^2$: Reduced $\chi^2$ from the SSP fitting.    
\end{tablenotes}  
\end{table}

A sample of the catalog is shown in Table~\ref{tab:catalog}. We provide the following properties of the AGNs:
\begin{itemize}
\item {\bf TARGETNAME}: 6dF ID, a combination of RA and Dec in sexagesimal units.
\item {\bf RA}: Right ascension in degrees.
\item {\bf DEC}: Declination in degrees.
\item {$\bf z$}: redshift value from \citet{6dF04, 6dFGS}.
\item {\bf Type}: AGN Type, T1=Type 1, K01=Type 2 satisfying \citet{Kewley01} criteria, K03 = Type 2 satisfying \citet{Kauffmann03} criteria.
\item {\bf SN$_V$}: Signal-to-noise ratio of the V-arm of the spectrum which contains the \Hb\ line. Since \Hb\ is the weakest of the four AGN identification lines, this is the measure of spectral quality which affects the ability to detect narrow emission lines for Type 2 AGN identification.
\item {\bf SN$_R$}: Signal-to-noise ratio of the R-arm of the spectrum which contains the \Ha\ line, the measure of spectral quality which affects the ability to detect the broad \Ha\ line for Type 1 AGN identification.
\item {\bf TMID}: 2MASS ID, a combination of RA and Dec in sexagesimal units, to mark the objects in common with ZCF \citep{ZCF19}.
\item {\bf SSP $\chi^2$}: The reduced $\chi^2$ from SSP fitting. We applied same SSP $\chi^2$ cut at 6.05 to keep 99\% of the successful fits as in ZCF \citep{ZCF19}.
\end{itemize}

\begin{table}
 \caption{Non-AGN Galaxies (Illustrative Extract)} 
 \label{tab:nonAGNs}
\tabletypesize{\scriptsize} 
\begin{tabular}{l r r l c l r r l l c} 
\hline\hline
  \multicolumn{1}{c}{TARGETNAME} &
  \multicolumn{1}{c}{RA \ (deg)} &
  \multicolumn{1}{c}{DEC\ (deg)} &
  \multicolumn{1}{c}{$z$} &
  \multicolumn{1}{c}{Type} &
  \multicolumn{1}{c}{TMID} &
  \multicolumn{1}{c}{SN$_V$} &
  \multicolumn{1}{c}{SN$_R$} &
  \multicolumn{1}{c}{PROB1} &
  \multicolumn{1}{c}{PROB2} &
  \multicolumn{1}{c}{PROBAGN} \\
\hline
  g2359411\_051614 & 359.92121 & -5.27053 & 0.117294 & -- &  & 6.0 & 10.2 & 0.0870 & 0.1811 & 0.2682\\
  g2359433\_540227 & 359.93050 & -54.04086 & 0.084917 & -- &  & 17.3 & 20.0 & 0.0646 & 0.1506 & 0.2152\\
  g2359470\_750411 & 359.94592 & -75.06958 & 0.058782 & Emi &  & 7.9 & 7.0 & 0.0956 & 0.1740 & 0.2696\\
  g2359485\_455710 & 359.95192 & -45.95289 & 0.065434 & -- &  & 8.4 & 7.5 & 0.0944 & 0.1727 & 0.2672\\
  g2359500\_205008 & 359.95829 & -20.83556 & 0.064346 & -- &  & 9.2 & 9.0 & 0.0910 & 0.1710 & 0.2620\\
  g2359553\_123752 & 359.98029 & -12.63114 & 0.104582 & -- &  & 6.1 & 10.1 & 0.0872 & 0.1808 & 0.2680\\
  g2359592\_691411 & 359.99675 & -69.23639 & 0.060147 & -- &  & 5.2 & 7.0 & 0.0945 & 0.1820 & 0.2765\\
  g2359390\_484804 & 359.91246 & -48.80111 & 0.065728 & Emi &  & 12.0 & 12.5 & 0.0831 & 0.1642 & 0.2473\\
  g2359437\_290936 & 359.93229 & -29.15992 & 0.027535 & Emi &  & 9.2 & 9.6 & 0.0896 & 0.1712 & 0.2609\\
  g2359547\_301207 & 359.97804 & -30.20192 & 0.030624 & -- &  & 10.7 & 10.3 & 0.0882 & 0.1672 & 0.2554\\
  g2359555\_392832 & 359.98104 & -39.47542 & 0.102470 & -- &  & 10.0 & 15.9 & 0.0735 & 0.1719 & 0.2454\\
  g0000086\_062226 & 0.03601 & -6.37399 & 0.021785 & Emi & 00000865-0622263 & 15.8 & 17.7 & 0.0705 & 0.1542 & 0.2247\\
  g0000236\_470108 & 0.09836 & -47.01881 & 0.019984 & -- & 00002363-4701076 & 23.0 & 10.3 & 0.0923 & 0.1277 & 0.2200\\
  g0000586\_333643 & 0.24409 & -33.61195 & 0.023063 & Emi & 00005858-3336429 & 13.9 & 13.5 & 0.0810 & 0.1587 & 0.2397\\
  g0001029\_431950 & 0.26204 & -43.33044 & 0.038783 & -- & 00010289-4319496 & 21.6 & 24.0 & 0.0544 & 0.1378 & 0.1922\\
  g0001175\_530035 & 0.32276 & -53.00967 & 0.032436 & Emi & 00011748-5300348 & 11.9 & 13.5 & 0.0804 & 0.1649 & 0.2453\\
  g0001330\_072610 & 0.38729 & -7.43614 & 0.029474 & Emi & 00013295-0726099 & 24.8 & 25.1 & 0.0518 & 0.1271 & 0.1789\\
\hline\end{tabular}
 \begin{tablenotes}
      \small
      \item TARGETNAME: 6dF ID; RA: right ascension; DEC: declination; $z$: redshift;  Type: galaxy type, `--' = Non-emission line galaxy, `Emi' = galaxy with Emission lines; TMID: Two MASS ID, indicate common galaxies from 2MRS; PROB1: Likelihood to be a Type 1 AGN,  i.e.,  $L_{sT1}$; PROB2: Likelihood to be a Type 2 AGN satisfying \citet{Kewley01} criteria,  i.e.,  $L_{sT2}$; PROBAGN: Likelihood to be an AGN,  sum of $L_{sT1}$ and $L_{sT2}$.      
 \end{tablenotes}  

\end{table}

\begin{table}
 \caption{The 6dF galaxies affected by the telluric contamination (Illustrative Extract)} 
 \label{tab:missing}
 \begin{tabular}{l r r l l l l c}  
\hline\hline
  \multicolumn{1}{c}{TARGETNAME} &
  \multicolumn{1}{c}{RA \ (deg)} &
  \multicolumn{1}{c}{DEC\ (deg)} &
  \multicolumn{1}{c}{$z$} &
  \multicolumn{1}{c}{TMID} &
  \multicolumn{1}{c}{PROB1} &
  \multicolumn{1}{c}{PROB2} &
  \multicolumn{1}{c}{PROBAGN} \\
\hline
  g0000009\_405412 & 0.00354 & -40.90330 & 0.049704 & 00000082-4054120 & 0.1084 & 0.1921 & 0.3005\\
  g0000190\_441241 & 0.07900 & -44.21150 & 0.048703 &  & 0.1084 & 0.1921 & 0.3005\\
  g0000202\_394843 & 0.08429 & -39.81197 & 0.042436 &  & 0.1084 & 0.1921 & 0.3005\\
  g0000251\_260240 & 0.10454 & -26.04450 & 0.050845 & 00002509-2602401 & 0.1084 & 0.1921 & 0.3005\\
  g0000328\_401323 & 0.13658 & -40.22292 & 0.043005 &  & 0.1084 & 0.1921 & 0.3005\\
  g0000358\_403432 & 0.14896 & -40.57561 & 0.050031 &  & 0.1084 & 0.1921 & 0.3005\\
  g0000399\_110927 & 0.16617 & -11.15750 & 0.045552 &  & 0.1084 & 0.1921 & 0.3005\\
  g0000444\_032854 & 0.18496 & -3.48172 & 0.047255 &  & 0.1084 & 0.1921 & 0.3005\\
  g0000459\_815803 & 0.19129 & -81.96742 & 0.042244 &  & 0.1084 & 0.1921 & 0.3005\\
  g0000495\_255356 & 0.20629 & -25.89875 & 0.051068 &  & 0.1084 & 0.1921 & 0.3005\\
  g0000523\_355037 & 0.21805 & -35.84368 & 0.051973 & 00005234-3550370 & 0.1084 & 0.1921 & 0.3005\\
  g0000532\_355911 & 0.22161 & -35.98630 & 0.050041 & 00005317-3559104 & 0.1084 & 0.1921 & 0.3005\\
  g0000555\_404245 & 0.23129 & -40.71239 & 0.048954 &  & 0.1084 & 0.1921 & 0.3005\\
  g0000558\_255421 & 0.23238 & -25.90581 & 0.050463 &  & 0.1084 & 0.1921 & 0.3005\\
  g0001007\_375430 & 0.25275 & -37.90836 & 0.050417 &  & 0.1084 & 0.1921 & 0.3005\\
  g0001080\_534149 & 0.28342 & -53.69697 & 0.048467 &  & 0.1084 & 0.1921 & 0.3005\\
  g0001083\_380735 & 0.28446 & -38.12636 & 0.045083 &  & 0.1084 & 0.1921 & 0.3005\\
  g0001094\_384406 & 0.28908 & -38.73511 & 0.051017 &  & 0.1084 & 0.1921 & 0.3005\\
  g0001224\_384521 & 0.34347 & -38.75589 & 0.051265 & 00012242-3845212 & 0.1084 & 0.1921 & 0.3005\\
\hline\end{tabular}
 \begin{tablenotes}
      \small
      \item Note: Due to the telluric contamination, these spectra are not usable for identifying AGN candidates. We therefore assigned the uniform AGN probability for these galaxies.
TARGETNAME: 6dF ID; RA: right ascension; DEC: declination; $z$: redshift; TMID: Two MASS ID, indicate common galaxies from 2MRS; PROB1: Likelihood to be a Type 1 AGN; PROB2: Likelihood to be a Type 2 AGN satisfying \citet{Kewley01} criteria; PROBAGN: Likelihood to be an AGN.  
 \end{tablenotes}  
\end{table}

In the online version of our catalog, we provide extended information for the identified AGNs, namely the FWHM and flux of the broad \Ha\ line, fluxes, and flux errors of the narrow lines used in AGN identification, and the $\rm (S/N)_{line}$ for each line. The additional information allows users  to create their own customized AGN catalog using different AGN selection criteria. Note that the flux values are not absolute fluxes due to the un-flux-calibrated spectra.   
Therefore, the reported fluxes should only be used as ratios.
AGNs with FWHM $ \geq$ 4000 \kms in broad \Ha\ line are likely to have underestimated FWHM values, as their broad lines are often lumpy and asymmetric which need multiple Gaussians to properly model the line profiles \citep[e.g., ][]{Greene07}.

\section{AGN DETECTION RATES}
\label{sec:agnrates}

Statistical correlation studies are an important use of AGN catalogs.
To fully understand the results of such studies, the completeness and homogeneity of the catalogs needs to be assessed. 
Our analysis in \citet{ZCF19} shows that the AGN detection rates are due to the overall continuum S/N of the spectrum. The line S/N is determined by the spectral quality (i.e., the continuum S/N) and the strength of the emission line in a spectrum. The continuum S/N is a measure of the noise at a given wavelength and the ability to detect a line at that wavelength.
Since data quality plays a crucial role in spectroscopically identifying AGNs, the AGN fraction is accordingly affected.

\begin{figure}[p]
\begin{center}
\includegraphics[width=0.45\textwidth,angle=0]{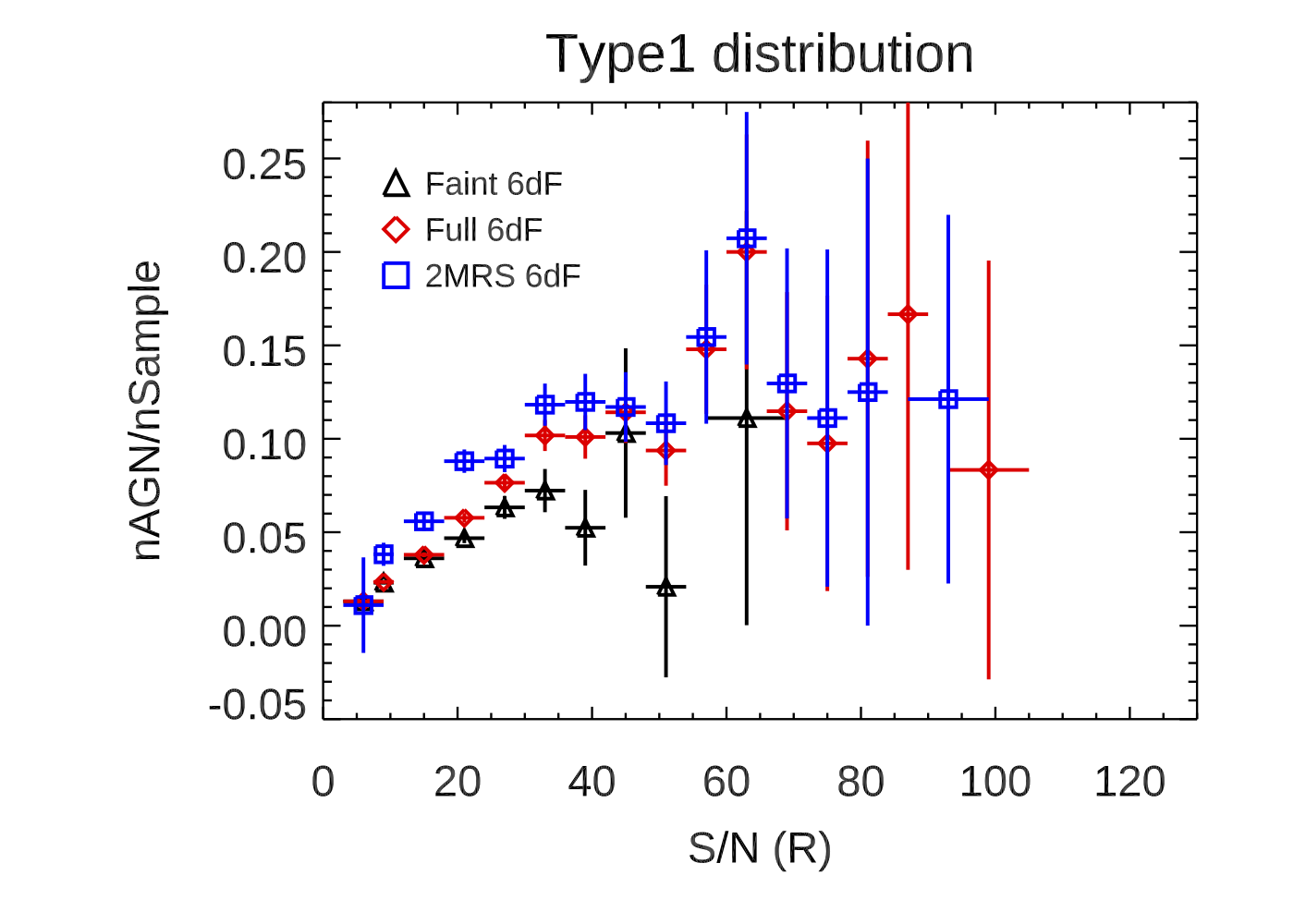}
\includegraphics[width=0.45\textwidth,angle=0]{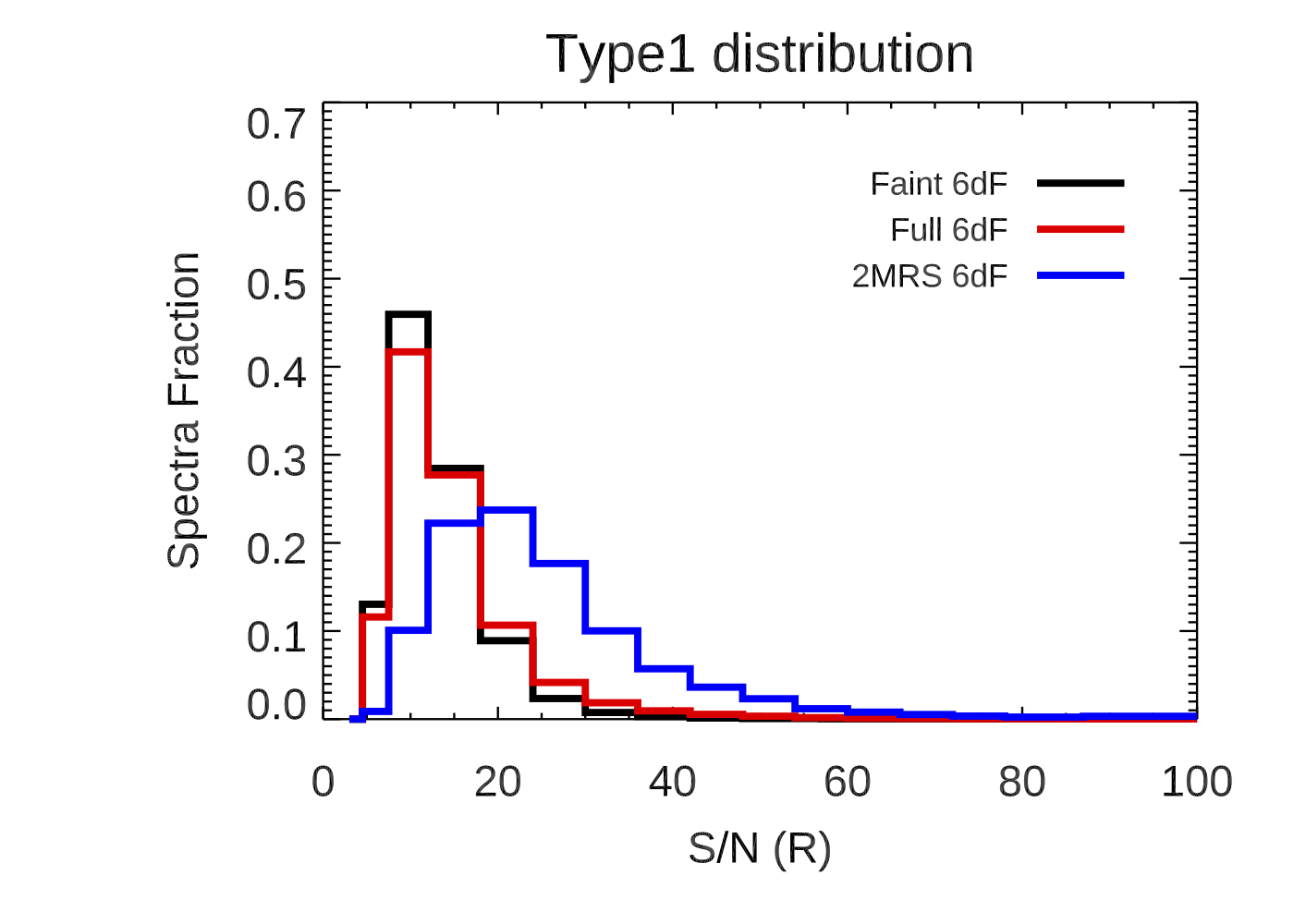}
\includegraphics[width=0.45\textwidth,angle=0]{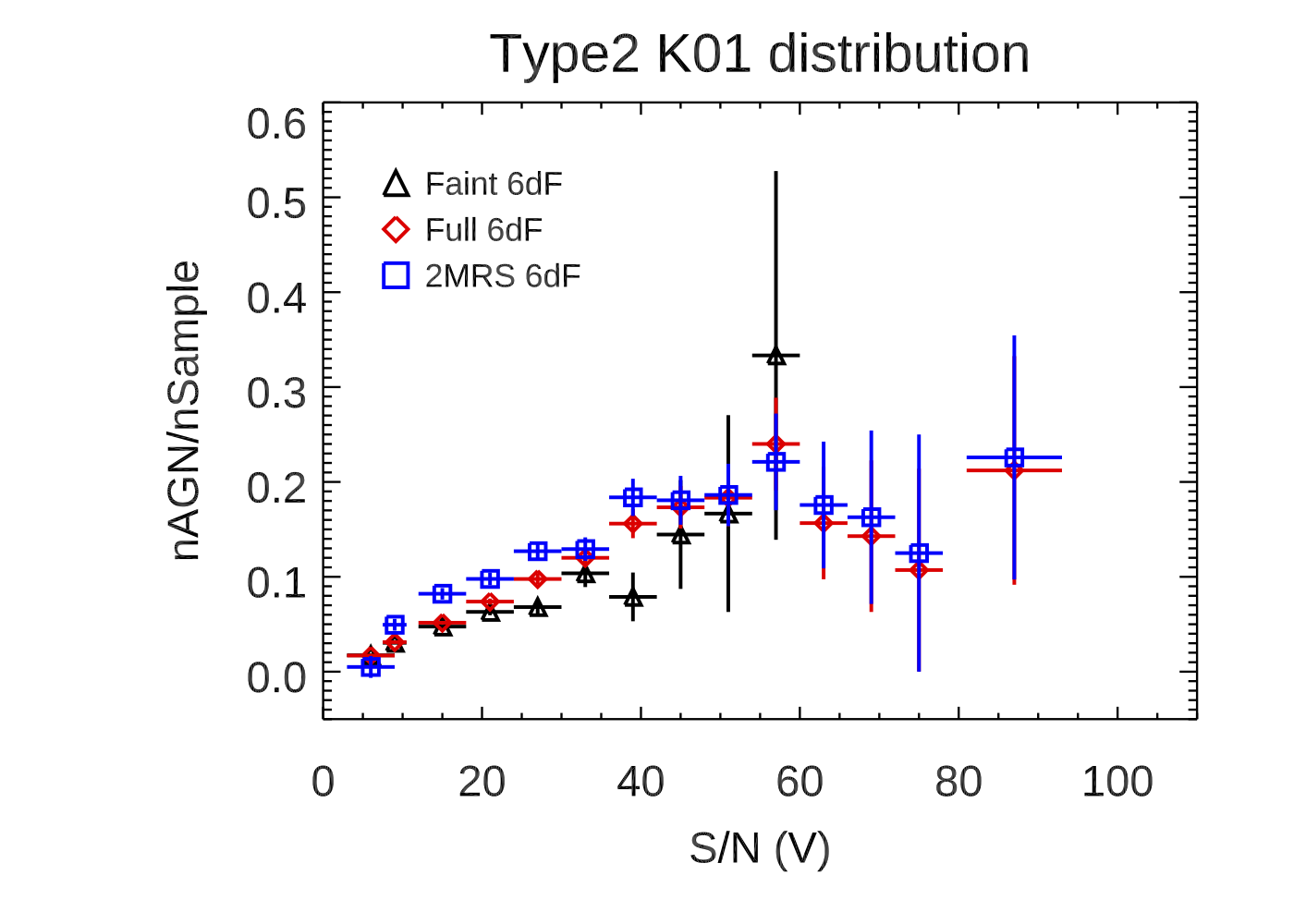}
\includegraphics[width=0.45\textwidth,angle=0]{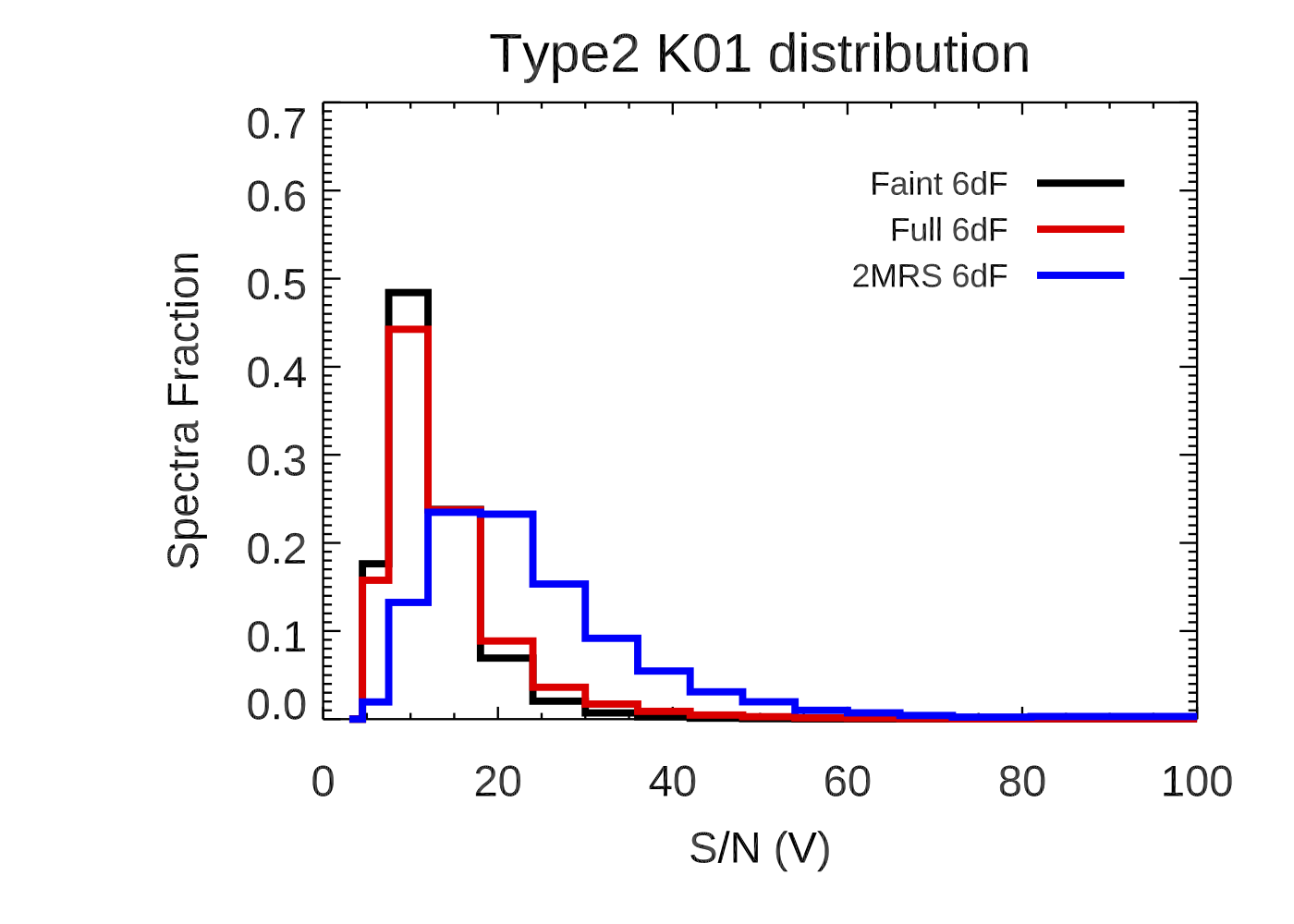}
\includegraphics[width=0.45\textwidth,angle=0]{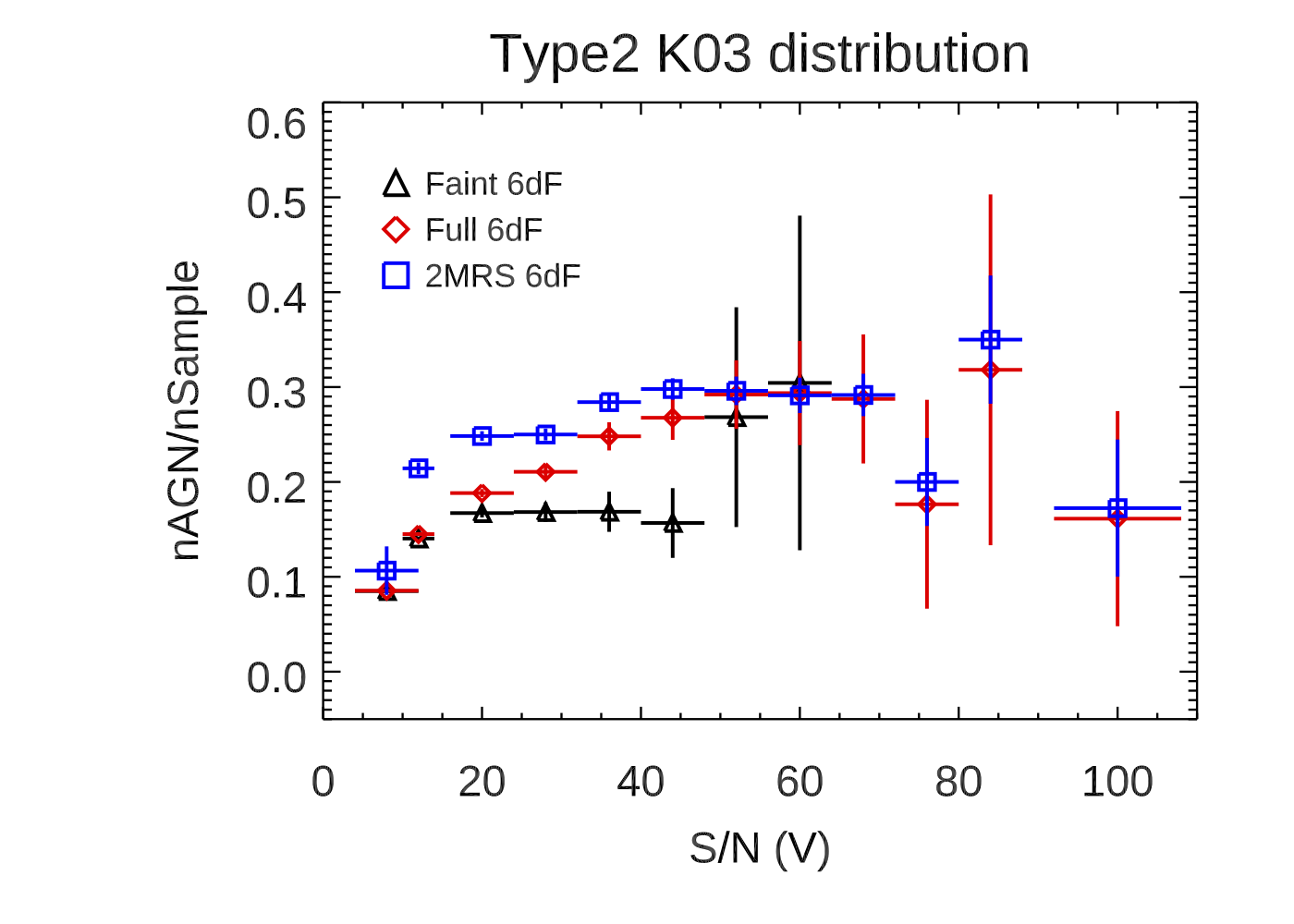}
\includegraphics[width=0.45\textwidth,angle=0]{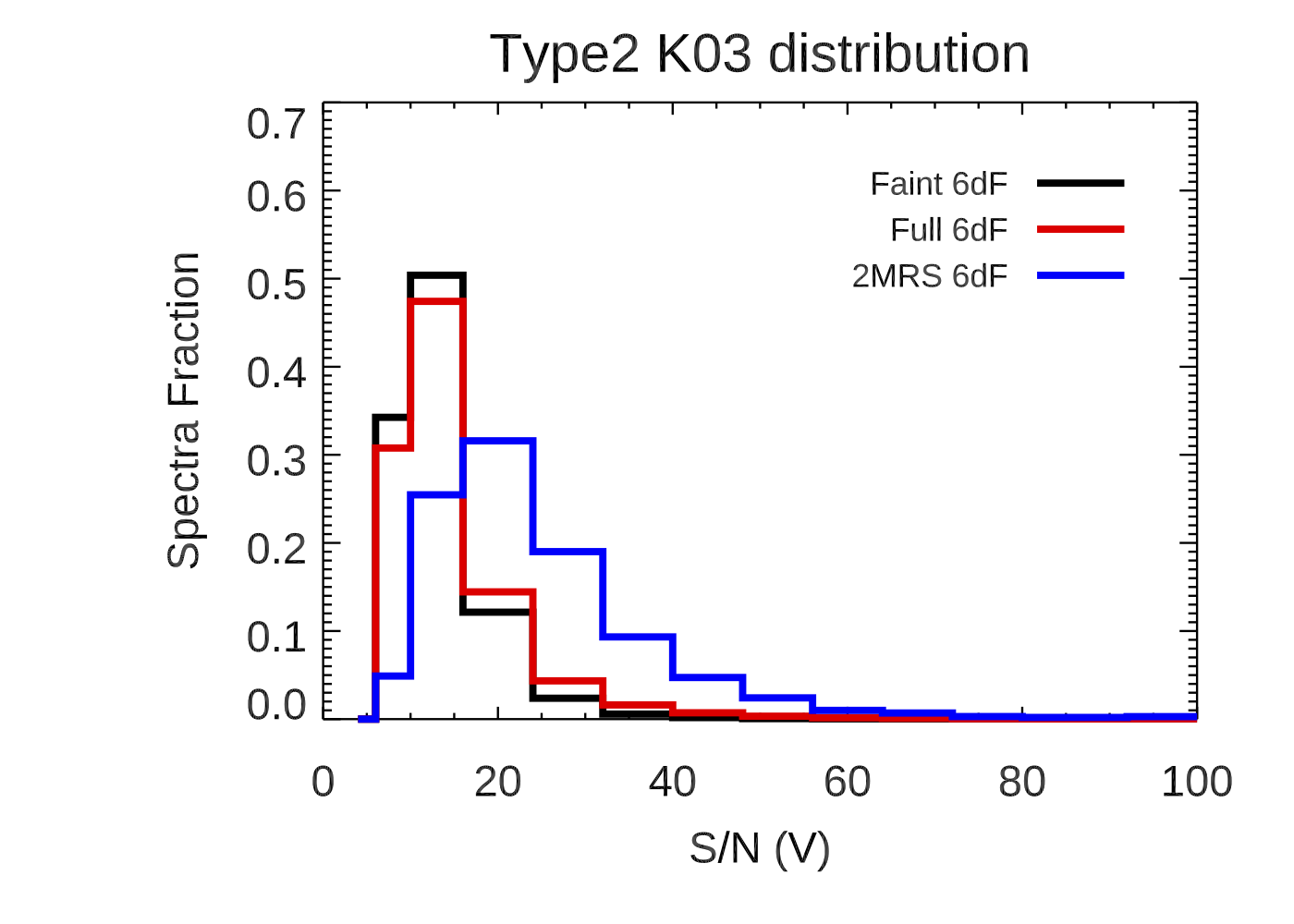}
\end{center}
\caption{AGN fraction (left column) and distribution (right column) as a function of the continuum S/N. S/N of R-band spectra is assessed for Type 1 AGN and S/N of V-band spectra is assessed for Type 2 AGN.
The top row is for Type 1 AGN, the middle row is for Type 2 AGN satisfying \citet{Kewley01} criteria, and the bottom row is for Type 2 AGN satisfying \citet{Kauffmann03} criteria. The AGN fraction increases with S/N and saturates at S/N $\gtrsim$ 40; beyond this value, the AGN fractions for the different subsamples are consistent with each other. } 
\label{fig:SN_AGN}
\end{figure}

As mentioned above, the 6dF data consist of two bands, namely V-band and R-band. The \Hb\  and [OIII] lines are normally located in the V-band spectra and the \Ha\ -[NII] lines are normally located in the R-band spectra. As the redshift of Full-6dF span a larger range, it is more robust to monitor the mean (S/N) of each bands rather than using a narrow range of spectra around \Hb\  and \Ha\  lines as we did in \citet{ZCF19}. The AGN detection rate as a function of continuum S/N is presented in Figure \ref{fig:SN_AGN}, where we also show the histograms of the S/N distributions.  As seen in the left column, the AGN detection rates increases with S/N for both the Full 6dF and the Faint 6dF, until the rate flattens at a saturation value of $S/N\sim 20-50$. The saturation S/N depends on the type of AGN. Type 1 AGN has a saturation value of $S/N\sim 40$, Type 2 AGN that satisfy the \citet{Kewley01} criteria has a saturation value of $S/N\sim 50$, and Type 2 AGN that satisfy with \citet{Kauffmann03} criteria has a saturation value of $S/N\sim 20$ in the Faint 6dF. Type 2 AGN that satisfy the \citet{Kauffmann03} criteria for the Full 6dF sample follows a smooth increasing curve, which is dominated by the 2MRS galaxies, while a more complicated behavior is seen in the Faint 6dF sample due to its lower S/N.
 As mentioned above, the Full 6dF has an overall lower S/N compared to the 2MRS subsample, a threshold is observed 
 for the AGN fraction, namely $S/N\sim 55$.  
 This behavior is consistent amongst the 2MRS 6dF, Faint 6dF and Full 6dF sample. 
 In the right column, we show the S/N distributions of the subsamples. As mentioned before, the 2MRS 6dF subsample has higher S/N on average.

\begin{table}
 \caption{AGN Numbers and Fractions}
 \label{tab:AGN}
 \begin{center}
\begin{tabular}{|l|l|r|r|c|}
\hline
  \multicolumn{1}{|c|}{} &
  \multicolumn{1}{c|}{Type 1} &
  \multicolumn{1}{c|}{Type 2 K01} &
  \multicolumn{1}{c|}{Type 2 K03} &
  \multicolumn{1}{c|}{N sample}  \\
\hline
  2MRS 6dF & 877 (8.46$\pm$0.30\%) & 1086 (10.47$\pm$0.33\%) & 2490 (24.01$\pm$0.54\%) & 10370\\ 
  Faint 6dF & 2332 (3.03$\pm$0.06\%) & 2779 (3.61$\pm$0.07\%) & 9666 (12.56$\pm$0.13\%) & 76953 \\
  Full 6dF& 3109 (3.56$\pm$0.06\%) & 3865 (4.43$\pm$0.07\%) & 12156 (13.92$\pm$0.13\%) & 87323\\
  \hline\end{tabular}
\end{center}
  \begin{tablenotes}
      \small
      \item The table lists the number of AGNs in each subsample as well as the AGN fraction and error on the fraction, relative to the number of spectra in our final sample, for Type 1 (broad line), Type 2 (narrow line) according to the \citet{Kewley01} (K01) criteria, and Type 2 according to the \citet{Kauffmann03} (K03) criteria. The AGN fractions are not the same for the different subsamples due to the differences in signal-to-noise, spectral resolution, and spectral sampling of each subsample.
    \end{tablenotes}
\end{table}

We show the AGN ratio across the sky (left panel) and its distribution as a function of redshift (right panel) in Figure~\ref{fig:AGNratio}. 
The inhomogeneity of AGN detection rates is observed in both panels, which is strongly affected by the data quality.
Data with lower redshift ($z < 0.07$) show higher AGN rates due to their higher data quality (as shown in Figure~\ref{fig:SN_AGN}). While data with higher redshifts tend to flatten out at an AGN detection rate around 0.15. The gap in redshift panel $z$ shows unavailable data due to telluric absorption, same effect is seen in Figure~\ref{fig:AGNz_vs_sn}. 

In order to be used in a rigorous statistical analysis, 
e.g., for correlations with the arrival direction of cosmic rays, we need to address the detailed 
inhomogeneity and incompleteness due to the missing or telluric-contaminated spectra. We perform similar corrections as \citet{ZCF19}.
We fit the AGN fraction as a function of the continuum S/N using a linear correlation below the saturation S/N value, and a constant above,
for Type 1 AGNs and Type 2 AGNs satisfying the \citet{Kewley01} criteria.\footnote{We use the \citet{Kewley01} criteria in this analysis because they yield a purer sample of galaxies with AGN activity. Furthermore, the AGN fraction for Type 2 AGNs satisfying the \citet{Kauffmann03} criteria seem to follow a curve rather than a line and requires a more complicated analysis.} Here are the relations we obtain from the fits:

\begin{alignat}{2}
R_{fT1} = & (SN1)*0.00261 -0.00128 & \; {\rm for} \, SN1 < 40 \nonumber\\  
             = & 0.108 \equiv R_{sT1} & \; {\rm for} \, SN1 \geq 40 \\
R_{fT2} =  & (SN2)*0.00346 -0.00184 & \; {\rm for} \, SN2 < 55 \nonumber\\  
             = & 0.192 \equiv R_{sT2} & \; {\rm for} \, SN2 \geq 55 
\end{alignat}

where $SN1$ ($SN2$) is the continuum signal-to-noise for R-band arm (V-band arm), i.e. containing the \Ha\ (\Hb) line, $R_{fT1}$ ($R_{fT2}$) is the AGN fraction for the given galaxy's S/N value for Type 1 (Type 2) AGNs, and $R_{sT1}$ ($R_{sT2}$) is the saturation value for the Type 1 (Type 2) AGN fraction. We assign a Type 1 AGN likelihood of zero for a non-AGN whose S/N (R-band arm) value is above 40. Similarly, we assign a Type 2 AGN likelihood of zero for a non-AGN whose S/N (V-band arm) is above 55. For galaxies with S/N values in the R-band and V-band arm less than the saturation values, we assign likelihoods as follows:
\begin{alignat}{2}
&L_{T1} && = \frac{R_{sT1}-R_{fT1}}{1-R_{fT1}-R_{fT2}}  \\
&L_{T2} && = \frac{R_{sT2}-R_{fT2}}{1-R_{fT1}-R_{fT2}}, 
\end{alignat}
where $L_{sT1}$ and $L_{sT2}$ are the likelihoods for the galaxy to be a Type 1 and Type 2 AGN, respectively. We assign the saturation AGN fractions as their AGN likelihoods for galaxies which are not in our spectroscopic sample. The probability for a galaxy to be an AGN is the sum of the likelihoods for Type 1 and Type 2 AGNs. In this work, those galaxies whose spectra contaminated by telluric features around the AGN diagnostic emission lines are not in our spectroscopic sample. We assign  them a flat likelihood of 0.3005, the saturation value for Type 1 and Type 2 AGNs.
The galaxies identified as AGNs are assigned an AGN likelihood of 1.0. 
Examples of the likelihood assignments for the non-AGNs and galaxies with spectra contaminated by telluric absorption are given in Tables~\ref{tab:nonAGNs} and \ref{tab:missing}, respectively. These likelihoods can be used to statistically correct for both the inhomogeneity and incompleteness. 
 These expressions for the likelihoods are only valid for the selection criteria used in this work, as specified in Sections \ref{sec:lineratios} and \ref{sec:catalog}. If the user choses different criteria, e.g. different line signal-to-noise ratio or BPT criteria, the homogeneity and completeness analyses should be repeated for the criteria used. 

We show the likelihood corrected AGN distribution, using the above equations (3) -- (6), across the sky and redshift in Figure~\ref{fig:AGNratio_whole}. The AGN rates are now homogeneously distributed.

\begin{figure}[thb]
\includegraphics[width=0.35\textwidth, angle=90]{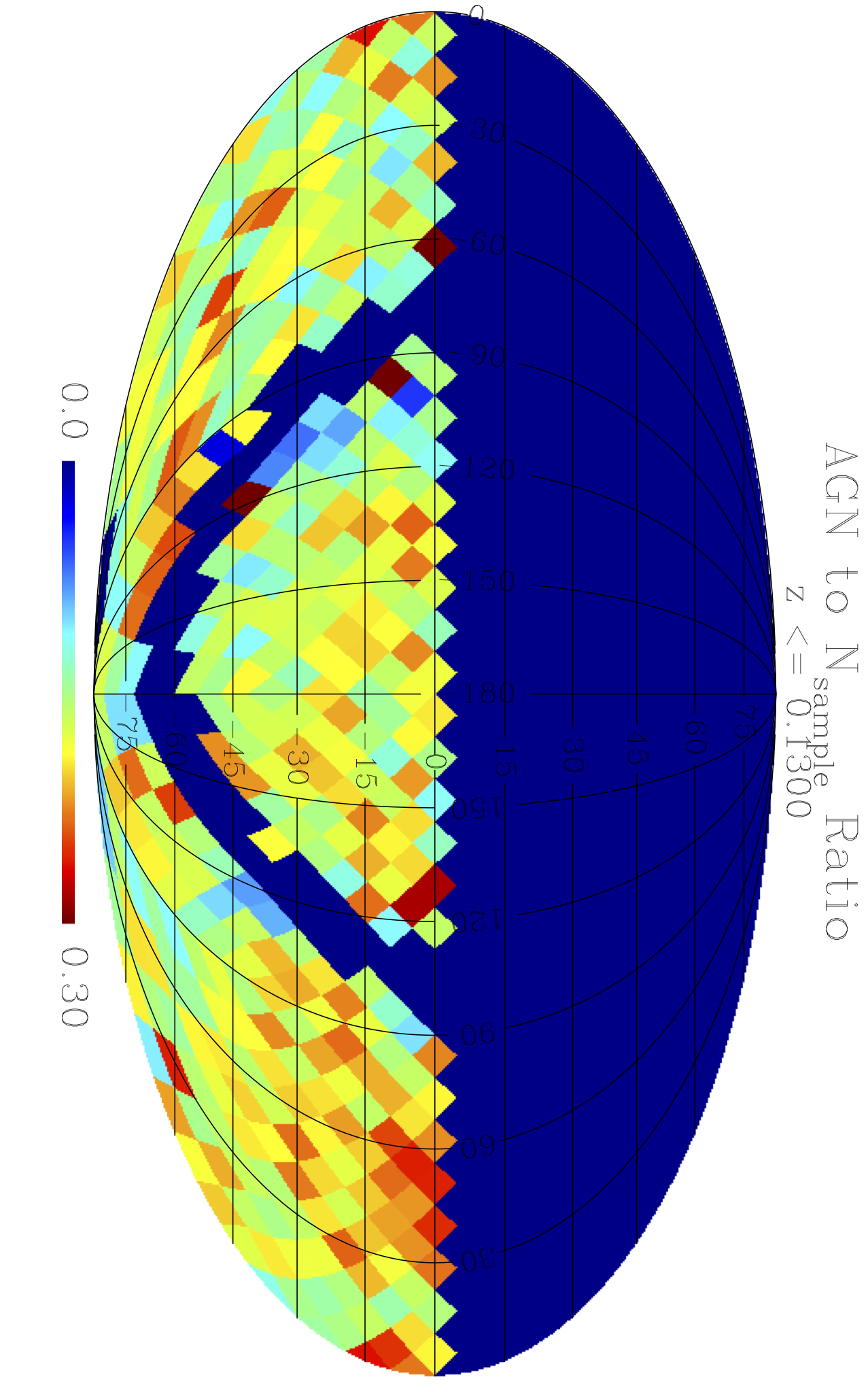} \includegraphics[width=0.45\textwidth, angle=0]{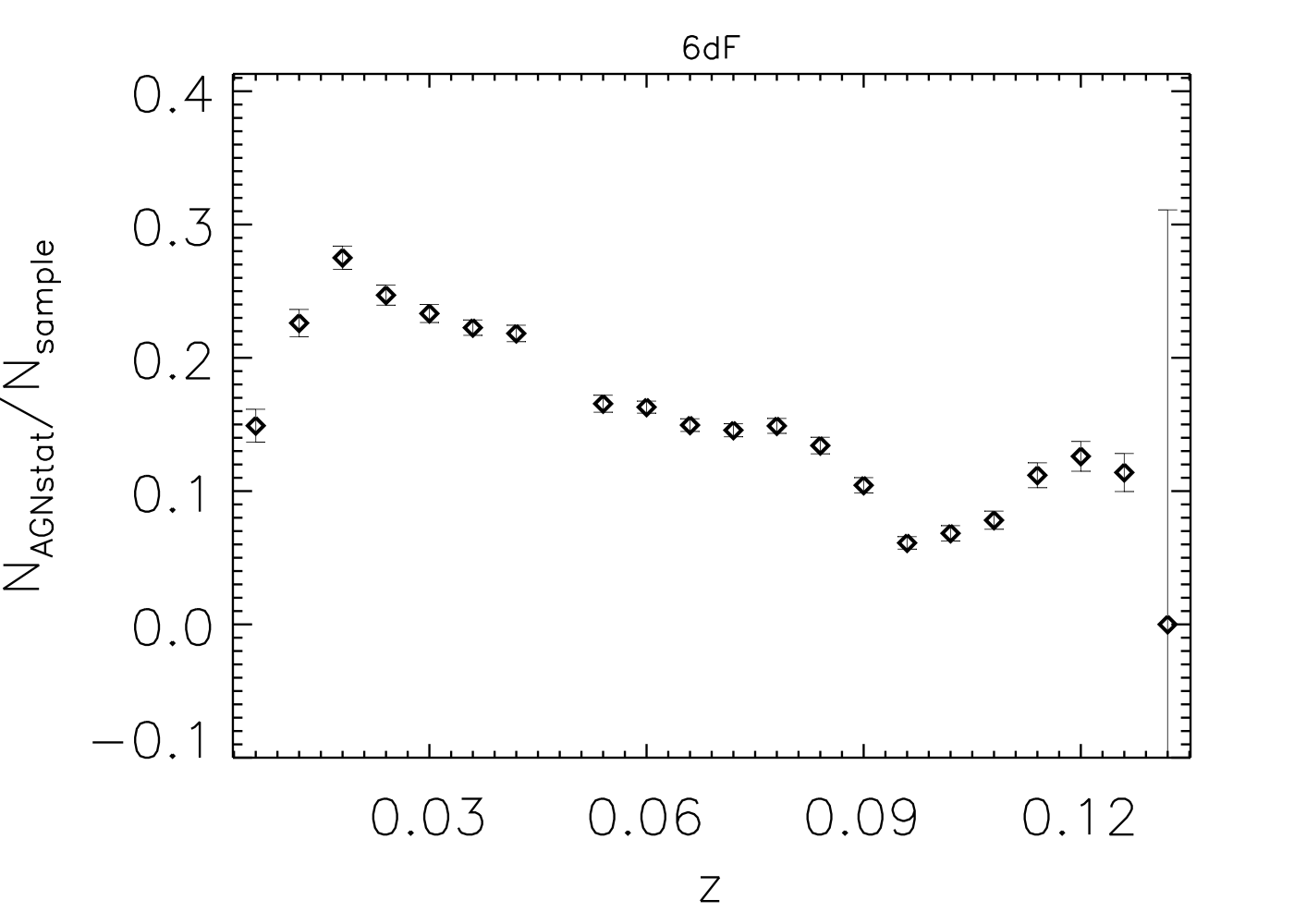} 
\caption{Left panel: 
Ratio of the number of identified AGNs to the number of galaxies in our spectroscopic sample ($N_{sample}$), across the Southern sky.
The sky has been divided into 768 equal-area regions, and the color indicates the AGN fraction.
Right panel: 6dF AGN fraction as a function of redshift.
In these plots, we have combined the broad line AGNs with the narrow line AGNs which satisfy the \citet{Kauffmann03} criteria.
$N_{sample}$ includes the spectra free from telluric contamination and a few hundred spectra with strong broad \Ha\ emission components that are affected by the telluric absorption.
}
\label{fig:AGNratio}
\end{figure}

\begin{figure}[thb]
\includegraphics[width=0.35\textwidth, angle=90]{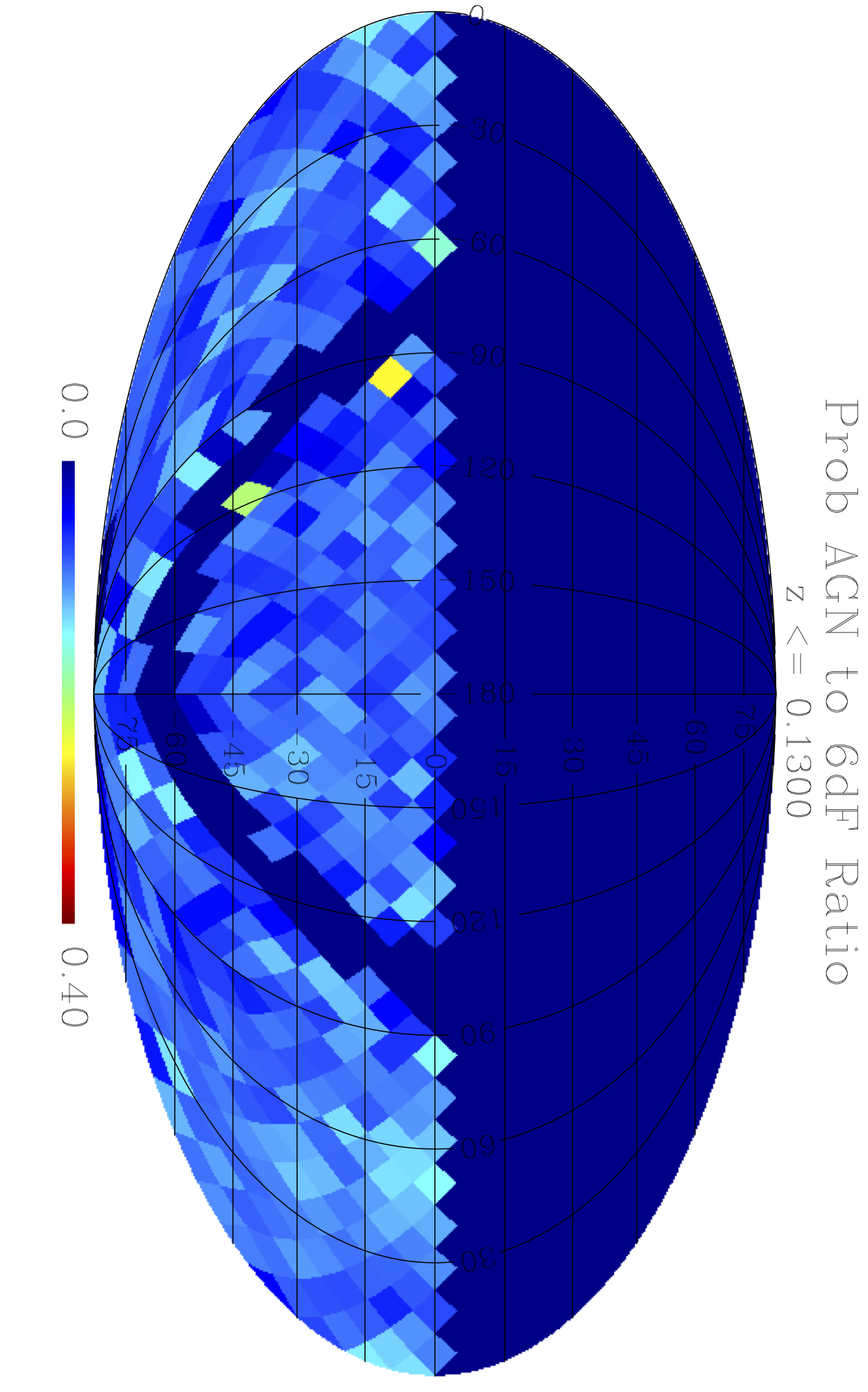}
\includegraphics[width=0.3\textwidth, angle=90]{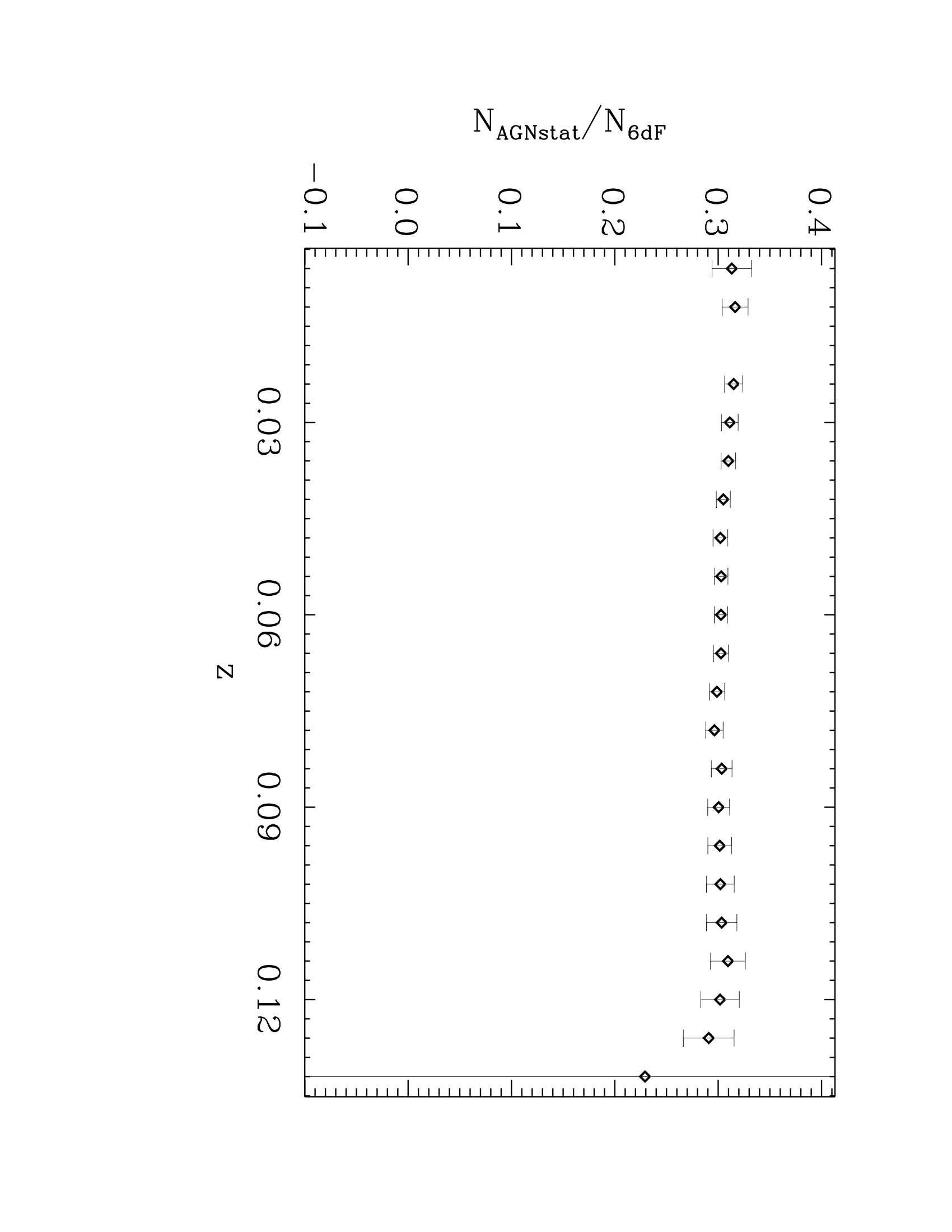}
\caption{Left panel: Ratio of the number of identified AGNs weighted by likelihood to the number of galaxies in 6dF, across the Southern sky.
The sky has been divided into 768 equal-area regions, and the color indicates the AGN fraction.
Right panel: Statistically weighted AGN fraction distribution as a function of redshift.
Homogeneous AGN fraction across the southern sky and in redshift is shown in the statistically weighted 6dF AGN catalog.
We have combined the broad line AGNs with the narrow line AGNs which satisfy the \citet{Kewley01} criteria in these plots.}
\label{fig:AGNratio_whole}
\end{figure}

\begin{figure}[thb]
\includegraphics[width=0.35\textwidth, angle=90]{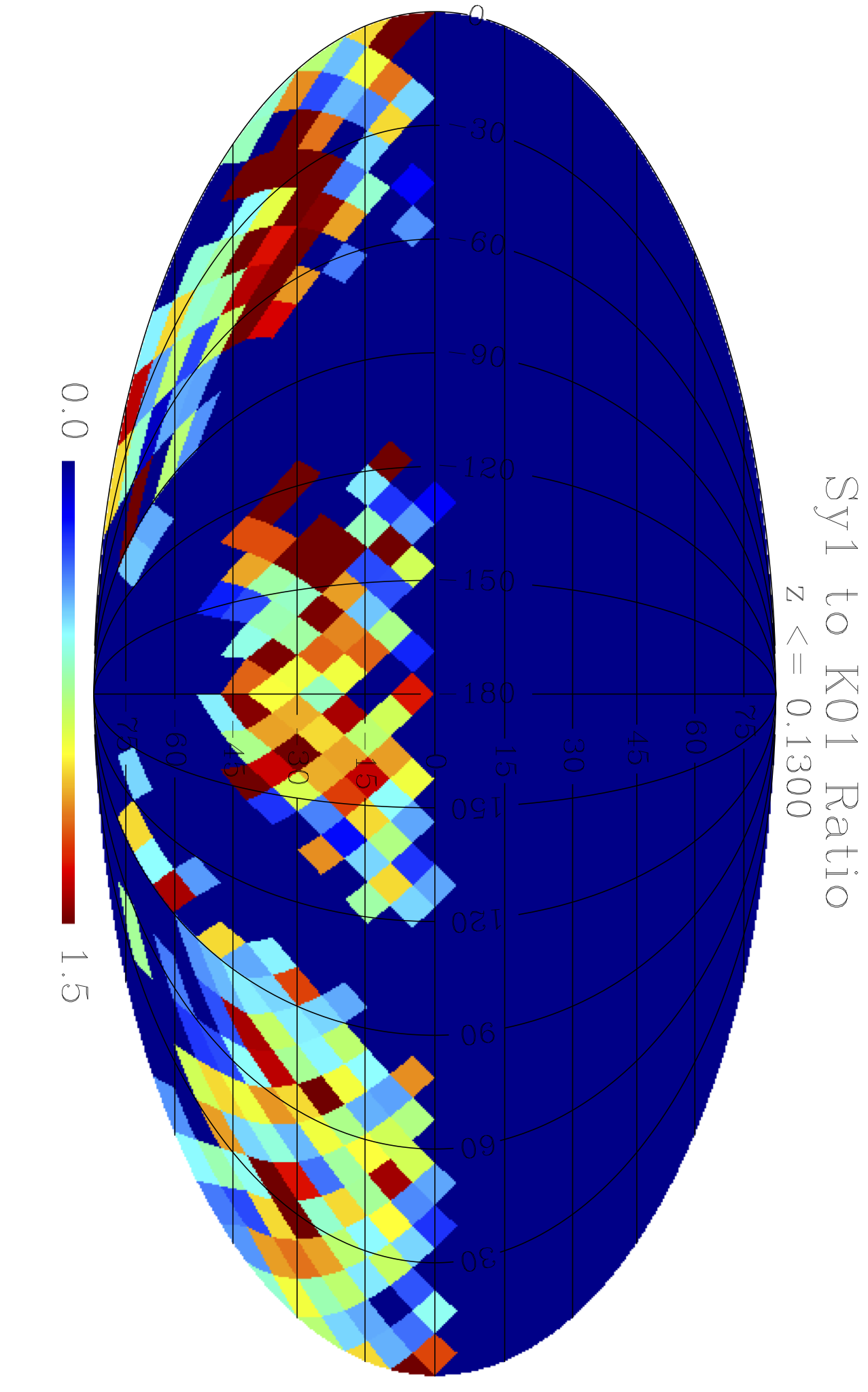}
\includegraphics[width=0.45\textwidth, angle=0]{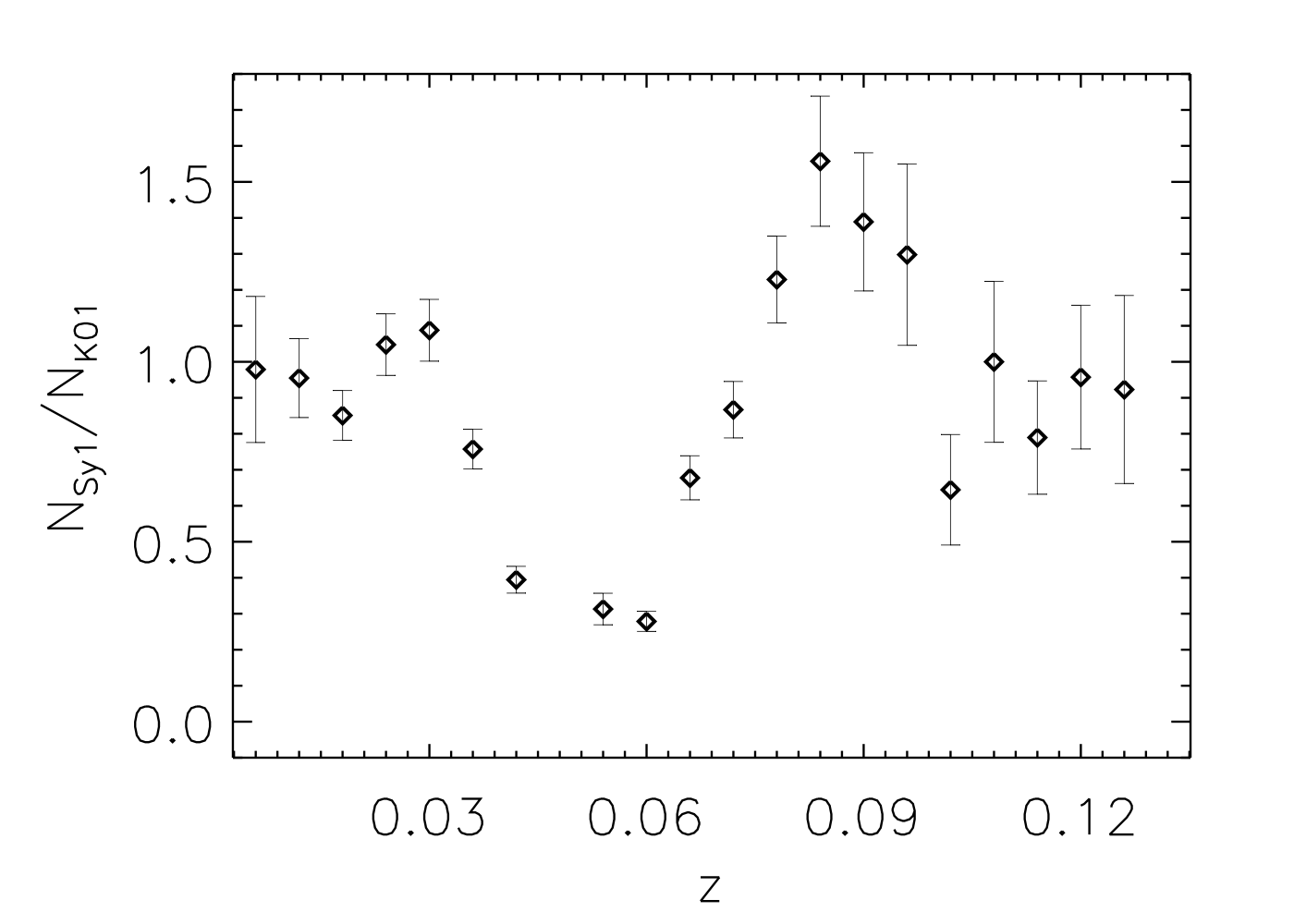}
\caption{Left panel: Ratio of the number of identified Type 1 (Sy1) AGNs  to the number of Type 2 (K01) AGNs in 6dF, across the Southern sky.
The sky has been divided into 768 equal-area regions, and the color indicates the AGN fraction.
Right panel:  Type 1 (Sy1) to Type 2 (K01) AGN fraction distribution as a function of redshift.}
\label{fig:sy1k01ratio_whole}
 \end{figure}

\begin{figure}[thb]
\includegraphics[width=0.35\textwidth, angle=90]{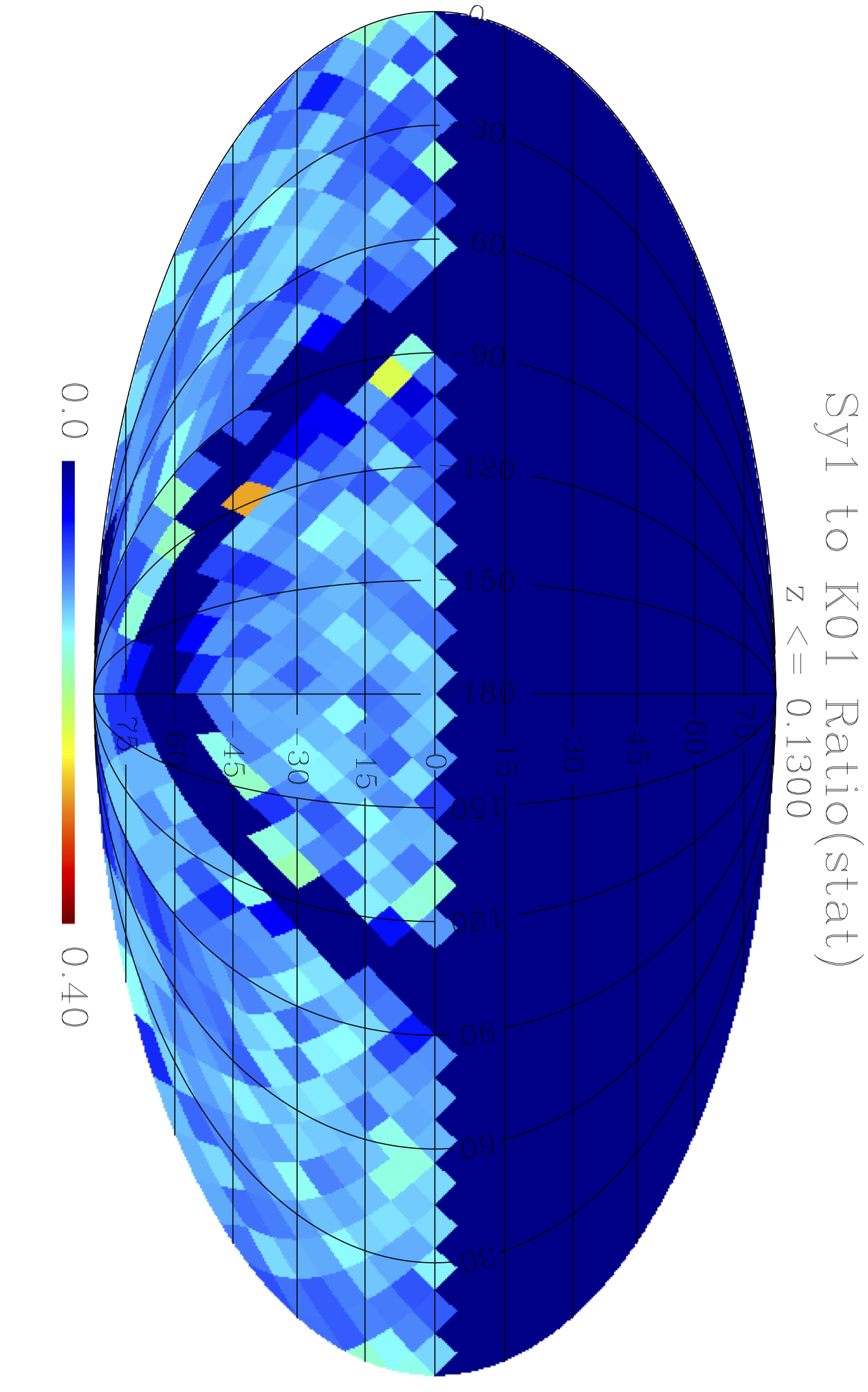}
\includegraphics[width=0.45\textwidth, angle=0]{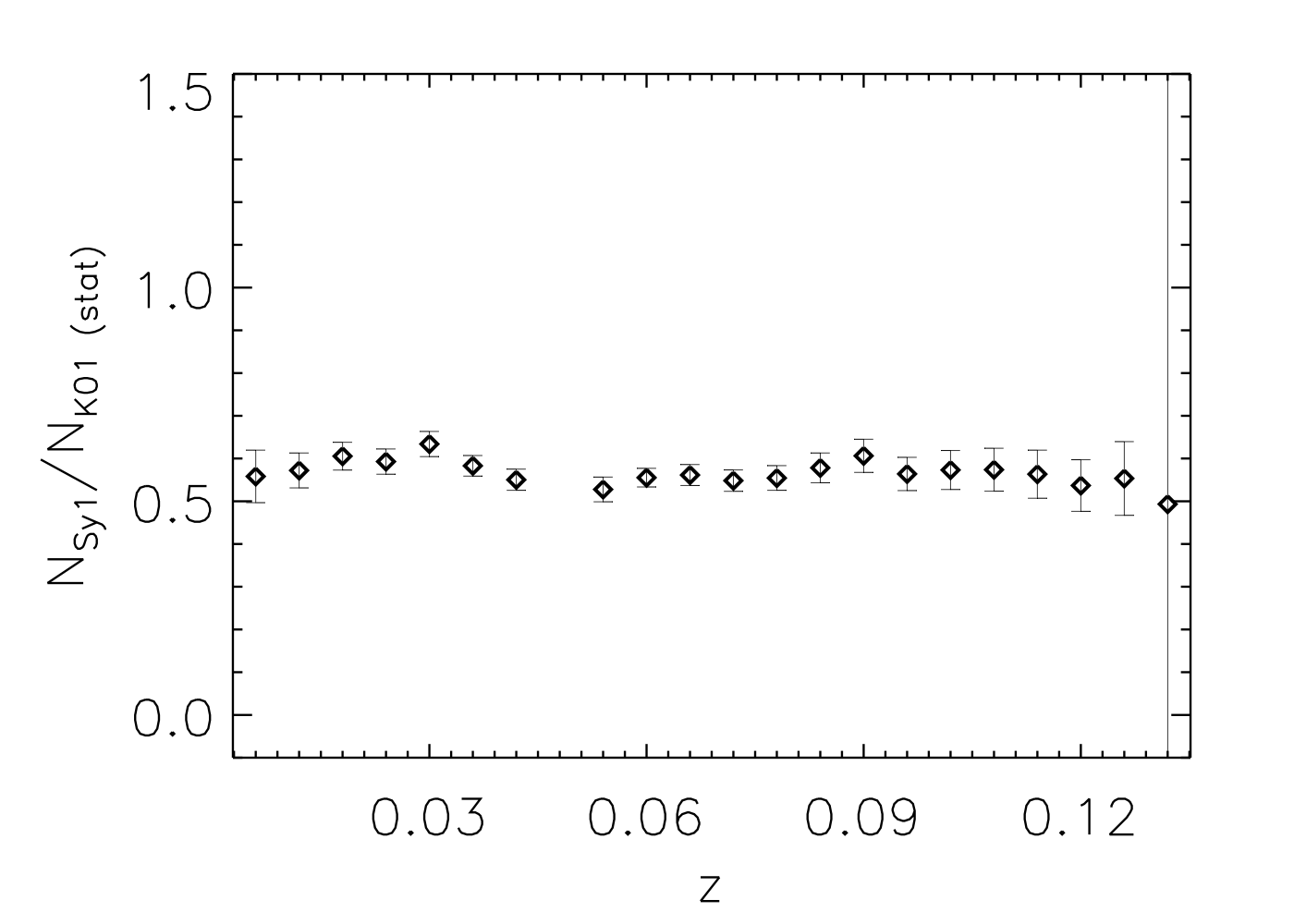}
\caption{Left panel: Ratio of the number of Type 1 (Sy1) AGNs  to the number of Type 2 (K01) AGN likelihood in 6dF
assigned based on continuum signal to noise across the Southern sky.
The sky has been divided into 768 equal-area regions, and the color indicates the AGN fraction.
Right panel: Statistically weighted Type 1 (Sy1) to Type 2 (K01) AGN fraction distribution as a function of redshift.
Nearly homogeneous Type 1 (Sy1)  to Type 2 (K01) AGN fraction across the southern sky and in redshift are seen in the statistically weighted 6dF AGN catalog.
}
\label{fig:sy1k01ratio_stat}
 \end{figure}

We also explore the  ratio of Type 1 (Sy1) to Type 2 (K01) AGN.  
The Type 1 to Type 2 AGN ratios vary widely across the sky and redshift as shown in Figure~\ref{fig:sy1k01ratio_whole}. 
This is expected since the Type 1 (Sy1) and Type 2 (K01) AGN fractions have different dependences on continuum signal-to-noise as seen in Figure~\ref{fig:SN_AGN}. 
We  correct the effect of signal-to-noise ($\rm S/N$) in assigning AGN likelihood by using equations (3) to (6), where Type 1 and Type 2 AGN likelihood are separately addressed. 
The statistical correction mostly eliminates the inhomogeneity in the Type 1/Type 2 ratio across the sky and in redshift, as shown in Figure~\ref{fig:sy1k01ratio_stat}.

This catalog, like other optical AGN catalogs, is affected by dust obscuration. The only indicator of reddening available to us is the Balmer decrement, H$\alpha$/H$\beta$, which has a nominal value of 3 for unobscured AGNs and increases with obscuration. However, due to lack of absolute flux calibration, and \Ha, \Hb\ located in two different arms, small values of Balmer decrements obtained from 6dF spectra are not reliable. Therefore, we cannot assess and correct for the inhomogeneity in this catalog due to dust obscuration.

\section{conclusions}
\label{sec:conclusions}

We have constructed a Southern-sky catalog of AGNs based on optical spectroscopy of the 6dF galaxy survey. 
The catalog consists of 3,109 broad line AGNs, and 12,156 narrow line AGNs which satisfy the \citet{Kauffmann03} criteria, of which 3,865 also satisfy the \citet{Kewley01} criteria. We report emission line widths, fluxes, flux errors, and signal-to-noise ratios for all the galaxies in our spectroscopic sample, allowing users to customize the selection criteria. We perform an assessment of the completeness and homogeneity of our catalog across the sky and in redshift. Due to the inhomogeneity in the S/N, the AGN fractions are not evenly distributed. Especially at higher redshift, data with lower S/N are dominating the sample, where the detection rates rapidly drop down. 
In order to minimize the effect of continuum signal-to-noise on Type 1 (Type 2) AGN detection efficiency, the minimum requirement for the continuum signal-to-noise ratio for a 6dF spectrum should be $\geq 55$ in R-band (S/N $\geq 40$ in V-band).

We provided means to correct 
for the S/N oriented inhomogeneity in our AGN catalog.  An AGN likelihood is assigned to the objects with low S/N and those whose diagnostic lines are contaminated by the telluric absorptions. After these corrections, the AGN fractions are uniform across the southern sky and in redshift, making the catalog suitable for rigorous statistical analysis.

\label{sec:conclu}
\acknowledgments

We thank the anonymous referee for his/her helpful comments.
We are grateful to Heath Jones and Philip Lah for their help with the 6dF spectra. 
We thank Joseph Gelfand for his helpful comments on our manuscript.
The research of GRF was supported by National Science Foundation grant NSF-PHY-2013199.
The research of I. Z. is supported by NYU Abu Dhabi Grant AD013.
Y. C. acknowledges the support of NYU Abu Dhabi Grant AD013 and NYU Abu Dhabi Grant AD022.

\end{document}